\documentclass[iop]{emulateapj}

\usepackage{natbib,aas_macros,amsmath}
\usepackage{multirow,color}
\usepackage{appendix}
\usepackage{comment}

\newcommand{\hi}{{\rm H{\sc i}}}
\newcommand{\lya}{Ly$\alpha$}
\newcommand{\lyb}{Ly$\beta$}
\newcommand{\delg}{$\delta_{\rm{gal}}$} 
\newcommand{\delF}{$\delta_{\langle F\rangle}$}
\newcommand{\Aobs}{$\rm{A_{obs}}$}
\newcommand{\Bobs}{$\rm{B_{obs}}$}
\newcommand{\Cobs}{$\rm{C_{obs}}$}
\newcommand{\Dobs}{$\rm{D_{obs}}$}
\newcommand{\Asim}{$\rm{A_{sim}}$}
\newcommand{\Bsim}{$\rm{B_{sim}}$}
\newcommand{\Csim}{$\rm{C_{sim}}$}
\newcommand{\Dsim}{$\rm{D_{sim}}$}

\begin{document}
\slugcomment{Accepted for publication in ApJ}
\shorttitle{Cosmic Galaxy-IGM \hi\ Relation at $\lowercase{z}\sim 2-3$}
\shortauthors{Mukae et al.}

\title{Cosmic Galaxy-IGM \hi\ Relation at $\lowercase{z}\sim 2-3$ Probed \\
in the COSMOS/UltraVISTA $1.6$ Deg$^2$ Field}

\author{
Shiro Mukae\altaffilmark{1,2},
Masami Ouchi\altaffilmark{1,3},
Koki Kakiichi\altaffilmark{4.5},
Nao Suzuki\altaffilmark{3},
Yoshiaki Ono\altaffilmark{1},
Zheng Cai\altaffilmark{6},\\
Akio K. Inoue\altaffilmark{7},
Yi-Kuan Chiang\altaffilmark{8},
Takatoshi Shibuya\altaffilmark{1},
Yuichi Matsuda\altaffilmark{9,10}}

\email{mukae@icrr.u-tokyo.ac.jp}
\altaffiltext{1}{
Institute for Cosmic Ray Research, The University of Tokyo, 5-1-5 Kashiwanoha, Kashiwa, Chiba 277-8582, Japan}
\altaffiltext{2}{
Department of Astronomy, Graduate School of Science, The University of Tokyo, 7-3-1 Hongo, Bunkyo, Tokyo, 113-0033, Japan}
\altaffiltext{3}{
Kavli Institute for the Physics and Mathematics of the Universe (Kavli IPMU, WPI), University of Tokyo, 5-1-5 Kashiwanoha, Kashiwa, Chiba, 277-8583, Japan}
\altaffiltext{4}{
Department of Physics and Astronomy, University College London, Gower Street, London WC1E 6BT, UK}
\altaffiltext{5}{
Max Planck Institute for Astrophysics, Karl-Schwarzschild-Strasse 1, D-85748 Garching bei M\"{u}nchen, Germany}
\altaffiltext{6}{
UCO/Lick Observatory, University of California, 1156 High Street, Santa Cruz, CA 95064, USA}
\altaffiltext{7}{
College of General Education, Osaka Sangyo University, 3-1-1, Nakagaito, Daito, Osaka 574-8530, Japan}
\altaffiltext{8}{
Department of Astronomy, University of Texas at Austin, 1 University Station C1400, Austin, TX 78712, USA}
\altaffiltext{9}{
National Astronomical Observatory of Japan, Osawa 2-21-1, Mitaka, Tokyo 181-8588, Japan}
\altaffiltext{10}{
Graduate University for Advanced Studies (SOKENDAI), Osawa 2-21-1, Mitaka, Tokyo 181-8588, Japan}

\begin{abstract}
We present spatial correlations of galaxies and IGM neutral hydrogen \hi\ in the COSMOS/UltraVISTA
$1.62$\ deg$^2$ field. Our data consist of 13,415 photo-$z$ galaxies at $z\sim 2-3$ with $K_s<23.4$ 
and the \lya\ forest absorption lines in the background quasar spectra selected from SDSS data
with no signature of damped \lya\ system contamination. We estimate a galaxy overdensity 
\delg\ in an impact parameter of $2.5$ (proper) Mpc, and calculate
the \lya\ forest fluctuations \delF\ whose negative values correspond to
the strong \lya\ forest absorption lines. 
We identify weak evidence of an anti-correlation
between \delg\ and \delF\ with a Spearman's rank correlation
coefficient of $-0.39$ suggesting that the galaxy overdensities and the \lya\ forest absorption lines
positively correlate in space at the $\sim90\%$ confidence level. This positive correlation indicates that
high-$z$ galaxies exist around an excess of \hi\ gas in the \lya\ forest. 
We find four cosmic volumes, dubbed \Aobs, \Bobs, \Cobs, and \Dobs,
that have extremely large (small) values of 
\delg\ $\simeq 0.8$ ($-1$) and  \delF\  $\simeq 0.1$ ($-0.4$),
three out of which, \Bobs--\Dobs, significantly depart from the \delg-\delF\ correlation,
and weaken the correlation signal.
We perform cosmological hydrodynamical simulations,
and compare with our observational results.
Our simulations reproduce the \delg -\delF\ correlation,
agreeing with the observational results.
Moreover, our simulations have model counterparts of \Aobs --\Dobs,
and suggest that the observations
pinpoint, by chance, a galaxy overdensity like a proto-cluster, gas filaments lying on the quasar sightline,
a large void, and orthogonal low-density filaments.
Our simulations  indicate that the significant departures 
of \Bobs--\Dobs\
are produced by the filamentary large-scale structures 
and the observation sightline effects.
\end{abstract}


\section{Introduction} \label{sec:Introduction}
The link between baryons and the cosmic web is a clue to understand both the galaxy formation and the baryonic processes in the large-scale structures (LSSs). 
The processes between galaxies and the intergalactic medium (IGM) are the inflow which represents gas accretion on to galaxies and the outflow driven by supernovae and active galactic nuclei. 
Neutral hydrogen \hi\ in the IGM is probed with the \lya\ forest absorption lines in spectra of background quasars \citep[e.g.,][]{Faucher-Giguere2008a, Becker2013a, Prochaska2013a} and bright star-forming galaxies \citep[e.g.,][]{Steidel2010a, Thomas2014a,  Mawatari2016a}. 

The detailed properties of galaxy-IGM \hi\ relations (hereafter galaxy-\hi\ relation) have been studied by spectroscopic observations of the  Keck Baryonic Structure Survey \citep[KBSS:][]{Rudie2012a, Rakic2012a, Turner2014a}, Very Large Telescope LBG Redshift Survey \citep[VLRS:][]{Crighton2011a, Tummuangpak2014a}, and other programs \citep[e.g.,][]{Adelberger2003a, Adelberger2005a}. 
These spectroscopic observations target \hi\ gas of the circumgalactic medium (CGM) around Lyman break galaxies (LBGs) that are high-$z$ star-forming galaxies identified with a bright UV and blue continuum. 

These LBG spectroscopy for the galaxy-\hi\ studies alone do not answer to the following two questions. 
One is the relation between IGM \hi\ and galaxies that are not selected as LBGs. 
Because LBGs are identified in their dust-poor star-forming phase, dust-rich and old-stellar population galaxies are missing in the past studies. 
In fact, the average star-formation duty cycle (DC) of LBGs is estimated to be $\sim30-60\%$ \citep{Lee2009a, Harikane2016a}. 
A large fraction of galaxies are not investigated in the studies of the galaxy-\hi\ relation. 
The other question is what the galaxy-\hi\ relation in a large-scale is. 
To date, the previous studies have investigated LBG-\hi\ relations around sightlines of background quasars within $\sim1$\ deg$^2$
corresponding to $\sim 70 \times 70$ comoving Mpc$^2$ at $z\sim2-3$ \citep{Adelberger2003a, Rudie2012a, Tummuangpak2014a}. 
There has been no study on the galaxy-\hi\ relation in a large-scale ($> 1$\ deg$^2$) at $z\sim2-3$. 
Only at $z \leq1$, \cite{Tejos2014a} conduct spectroscopic surveys to investigate galaxy-\hi\ relations in a large-scale ($> 1$\ deg$^2$). \cite{Tejos2014a} present the clustering analysis of spectroscopic galaxies and \hi\ absorption line systems.
At $z\sim2-3$, \cite{Cai2016a} have studied the galaxy-\hi\ relation focusing on extremely massive overdensities with $\sim6, 000$ SDSS quasar spectra by the MApping the Most Massive Overdensity Through Hydrogen (MAMMOTH) survey, but the galaxy-\hi\ relations have not been systematically explored.

We investigate spatial correlations of $K_s$-band selected galaxies with no DC dependence and IGM \hi\ at $z \sim 2-3$ in a large $1.62$\ deg$^2$ area of COSMOS/UltraVISTA field, in conjunction with the comparisons with our models of the cosmological hydrodynamical simulations.  We probe one large field contiguously covering LSSs.
Our study of the galaxy-\hi\ spatial correlation is complementary to the on-going programs of the MAMMOTH and the \lya\ forest tomography survey of the COSMOS Lyman-Alpha Mapping And Tomography Observations \citep[CLAMATO:][]{Lee2014b, Lee2016a} which aims at illustrating the distribution of IGM \hi\ gas in LSSs. In contrast, our study focuses on a spatial relation between galaxies and IGM \hi\ gas.

This paper is organized as follows. We describe the details of our sample galaxies and background quasars in Section \ref{sec:DATA}. In Section \ref{sec:Galaxy overdensity and HI absorption}, our data analysis is presented. We investigate the galaxy-\hi\ relation based on the observational data in Section \ref{sec:Galaxy-HI Correlation}. We introduce our simulations to examine the galaxy-\hi\ relation of our observational results in Section \ref{sec:SIMULATIONS}. In Section \ref{sec:DISCUSSION}, we compare observation and simulation results, and interpret our observational findings. Finally, we summarize our results in Section \ref{sec:SUMMARY}. 

Throughout this paper, we adopt  AB magnitudes \citep{Oke1983a}. We use a cosmological parameter set of $( \Omega_m, \Omega_\Lambda, \Omega_b, \sigma_8, n_s, h) = $ $( 0.26, 0.74, 0.045, 0.85, 0.95, 0.72)$ consistent with the nine-year $WMAP$ result \citep{Hinshaw2013a}.
We denote pkpc and pMpc (ckpc and cMpc) to indicate distances in proper (comoving) units.

\section{DATA} \label{sec:DATA}
\subsection{Photometric Galaxy Samples} \label{subsec:Photometric Galaxy Samples}
We investigate galaxy overdensities in the COSMOS/UltraVISTA field.  
Our photometric galaxy sample is taken from the COSMOS/UltraVISTA catalog that is a $K_s$-band selected galaxy catalog \citep{Muzzin2013a} 
made in the $1.62$\ deg$^2$ area of UltaVISTA DR1 imaging region \citep{McCracken2012a} in the COSMOS field \citep{Scoville2007a}. 

The COSMOS/UltraVISTA catalog consists of point-spread function matched photometry of 30 photometric bands. 
These photometric bands cover the wavelength range of 0.15-24 ${\rm \mu m}$ that includes the $GALEX$ FUV and NUV \citep{Martin2005a}, Subaru/SurimeCam \citep{Taniguchi2007a}, CFHT/MegaCam \citep{Capak2007a}, UltraVISTA \citep{McCracken2012a}, and $Spitzer$ IRAC+MIPS data \citep{Sanders2007a}. 
Photometric redshifts for all galaxies are computed with the EAZY code \citep{Brammer2008a}. 
The catalog contains $\sim150,000$ galaxies at $z\sim0-5$.
Stellar masses are determined by SED fitting by the FAST code \citep{Kriek2009a} with stellar population synthesis models.

We use the criteria of \cite{Chiang2014a} to select $z\sim 2-3$ photo-$z$ galaxies from the catalog.
We apply a $90\%$ completeness limit of $K_s <23.4$ mag that corresponds to a stellar mass limit of $\log_{10}(M_*/ M_{\odot}) > 9.7$ at $z\sim2.5$.
Note that this stellar mass limit of $z=2.5$ differs only by $3\%$ at the edges of our redshift window, $z=2$ and $3$.
This stellar mass limit is as large as $\sim0.1M^*$ at $z\sim2.5$ where $M^*$ is the characteristic stellar mass of a Schechter function parameter \citep{Schechter1976a} for the stellar mass functions (SMFs) taken from \cite{Muzzin2013b}.
We remove objects ($\sim4\%$) whose photometric redshifts show broad and/or multi-modal redshift probability distributions indicating poorly determined redshifts. 
Here, the photo-$z$ galaxies that we use have the redshift distribution function whose $80$ percentileprobability distribution extend no larger than $\Delta z = 0.2$ from the best estimate redshift.
Finally, our photometric samples consist of 13,415 photo-$z$ galaxies at $z\sim 2-3$ with $K_s<23.4$.

\subsection{Background Quasar Samples} \label{subsec:Background Quasar Samples}
We search for the \lya\ forest absorption lines found in background quasar spectra in the COSMOS/UltraVISTA field.
Our background quasar spectra are primarily taken from the BOSS Data Release 9 (DR9) Lyman-alpha Forest Catalog\ \citep[][hereafter L13]{Lee2013a}. 
L13 has reproduced quasar continua by the technique of mean-flux regulated principal component analysis (MF-PCA) continuum fitting \citep{Lee2012a}.
Because L13 does not include all quasars identified by the SDSS-III surveys \citep{Eisenstein2011a}, our background quasar spectra are also taken from the BOSS Data Release 12 (BOSS DR12) and the SDSS-III Data Release 12 (SDSS DR12) \citep{Alam2015a}. The BOSS DR9 and DR12 spectra are covered in the wavelength range of 3600-10400\AA. The SDSS DR12 spectra are obtained in the wavelength range of 3800-9200\AA \  that is slightly narrower than the BOSS wavelength range. Both the BOSS and the SDSS spectra have the spectral resolution of $R\equiv \lambda/ \Delta \lambda \approx 2000$.  

Because L13 compile spectra of quasars with a redshift range of $z_{{\rm qso}}> 2.15$, we search for background quasars at $z_{{\rm qso}}> 2.15$ from the BOSS and the SDSS data.
We find a total of 26  background quasars in the COSMOS/UltraVISTA field. We remove 4 background quasars that are located at the edge of the COSMOS/UltraVISTA field, because the cylinder volumes of these background quasars are cut by the COSMOS/UltraVISTA-field border by $>50$\%. 

To identify the \lya\ forest absorption lines, we adopt the \lya\ forest wavelength range of $1041-1185\rm{\AA}$ in the quasar rest frame.
With the speed of light $c$,
this wavelength range is defined as
\begin{equation}
 {\lambda_{\rm Ly\beta}} \left( 1 +\frac{5000\ \rm km\, s^{-1}}{c} \right)
< \lambda <
 {\lambda_{\rm Ly\alpha}} \left( 1 -\frac{8000\ \rm km\, s^{-1}}{c} \right),
\end{equation}
where $\lambda_{\rm Ly\beta}$ and $\lambda_{\rm Ly\alpha}$ are the rest-frame  wavelengths of the hydrogen Ly$\beta$ (1025.72\AA) and Ly$\alpha$ (1215.67\AA) lines, respectively.
Here, we include the velocity offsets of 5000 and 8000 km s$^{-1}$,
avoiding the \lyb\ forest contamination and the quasar proximity effect, respectively.

In the observed spectra, the \lya\ forest wavelength range shifts by a factor of $(1+z_{{\rm qso}})$. 
We search for background quasars whose \lya\ forest wavelength ranges cover \lya\ absorption lines. Because we investigate \lya\ absorption lines at $2\leq z{_{\rm{Ly\alpha}}} \leq 3$ of the COSMOS/UltraVISTA field, \lya\ absorption wavelength range is $3647{\rm{\AA}} \leq \lambda_{\rm{Ly\alpha}}(1+z{_{\rm{Ly\alpha}}}) \leq 4862{\rm{\AA}}$ in the observed frame that requires $2.08 \leq z_{{\rm qso}} \leq 3.67$. 
Limiting $z_{{\rm qso}}> 2.15$ given by the L13 spectra,
we select 21 background quasars at $2.15 < z_{{\rm qso}} \leq 3.67$ further removing 1 spectrum at $z_{{\rm qso}} > 3.67$.

Then, we investigate qualities of background quasar spectra. 
We define $\rm{S/N_{Ly\alpha}}$ as the median signal-to-noise ratio ($\rm{S/N}$) per pixel 
over the \lya\ forest wavelength range ($1041-1185\rm{\AA}$).
Because we find that the absorption signals are not reasonably obtained 
in 7 background quasar spectra with $\rm{S/N_{Ly\alpha}}<2$ by visual inspection,
we remove these 7 quasar spectra in our analysis.

We discard 4 broad absorption line (BAL) quasar spectra referring to the SDSS database. In addition, we check background quasar spectra by visual inspection, and remove 1 spectra with large flux fluctuations originated from unknown systematics. 

\begin{figure}[t]
	\centering
	\includegraphics[clip,bb=5 150 600 700,width=1.0\hsize]{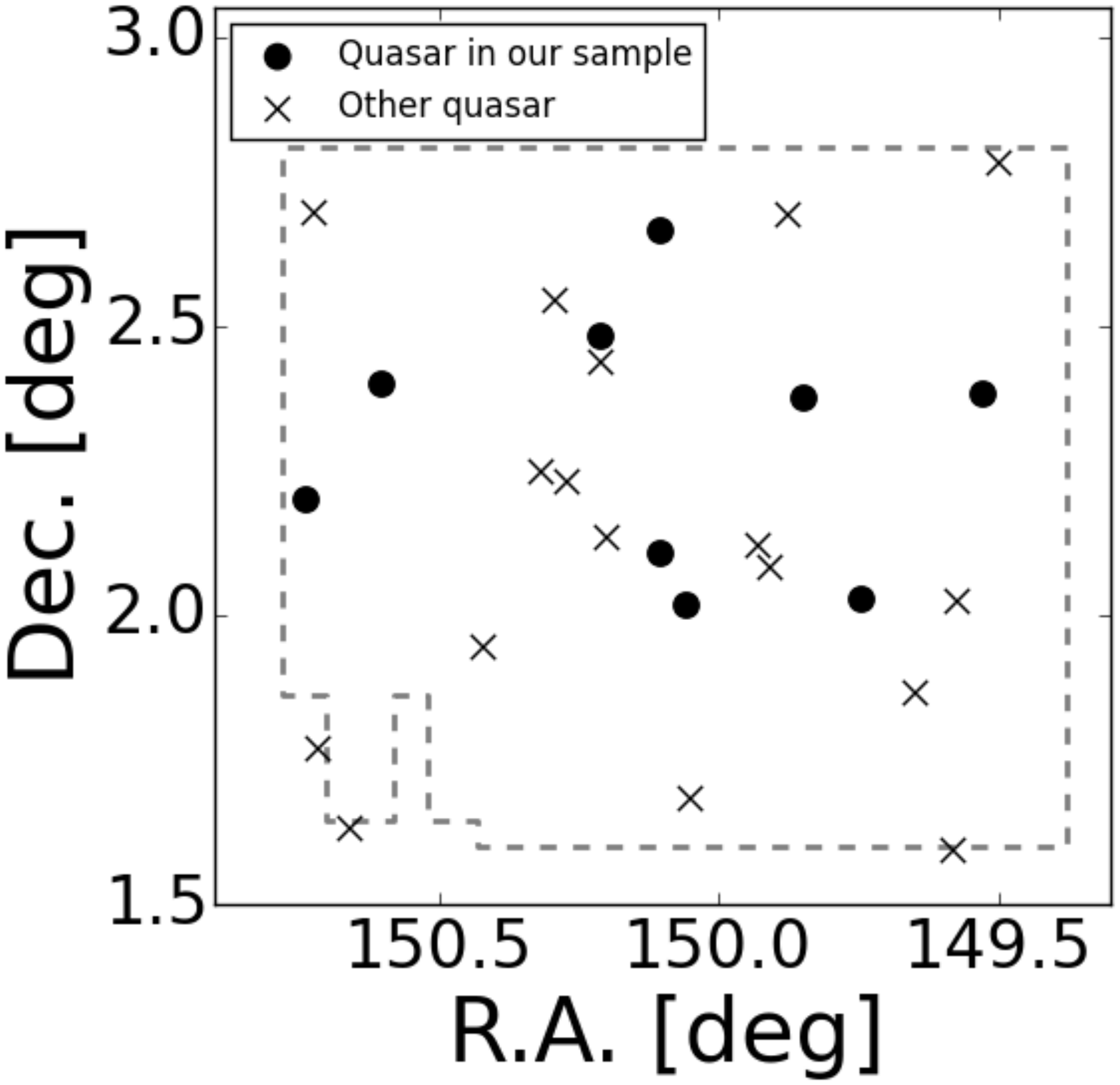}
	\caption{
	Sky distribution of the 26 background quasars found in the COSMOS/UltraVISTA field.
	The circles indicate background quasars at $z_{{\rm qso}} > 2.15$ analyzed in our study,
	while the crosses are those removed from our analysis.
	The gray dashed lines show the coverage of the COSMOS/UltraVISTA.
	}
	\label{fig:bkgqso_distribution}
\end{figure}

Finally, we use 9  background quasar spectra in the COSMOS/UltraVISTA field. 
Figure \ref{fig:bkgqso_distribution} shows the distribution of the background quasars in the COSMOS/UltraVISTA field. 
The mean $\rm{S/N_{Ly\alpha}}$ in our quasar samples is $\rm{S/N_{Ly\alpha}}\sim5$.

\section{Galaxy overdensity and HI absorption} \label{sec:Galaxy overdensity and HI absorption}
\subsection{Galaxy Overdensity} \label{subsec:Galaxy Overdensity}
We estimate galaxy overdensities around the quasar sightlines where the \lya\ forest absorption lines are observed.
The galaxy overdensities are calculated with the COSMOS/UltraVISTA catalog (\S\ \ref{subsec:Photometric Galaxy Samples}).
The galaxy overdensity $\delta_{\rm{gal}}$ is defined as
\begin{eqnarray}
	\delta_{\rm{gal}} \equiv \frac{n_{\rm{gal}}(z)}{\overline{n}_{\rm{gal}}(z)}-1,
\end{eqnarray}
where $n_{\rm{gal}}$ ($\overline{n}_{\rm{gal}}$) is the galaxy (average) number density in a cylinder at the redshift $z$ of the cylinder center. 
The redshift range of $n_{\rm{gal}}$ and $\overline{n}_{\rm{gal}}$ are defined by the redshift range of the cylinder length.
The base area of the cylinder is defined by a radius of $r=5'$ corresponding to an impact parameter of $\sim 2.5$ pMpc at $z\sim2.5$.  The length of the cylinder along a line of sight is given by an average photometric redshift uncertainty that corresponds to $25$ pMpc.
The estimated average photometric redshift uncertainty is $\sigma_{z}= 0.025(1+z)$ for galaxies with $K_s <23.4$ mag at $2\leq z \leq 3$ \citep{Scoville2013a}. The error of \delg\ is estimated with the combination of the photo-$z$ uncertainties and the Poisson errors.

\subsection{\lya\ forest absorption lines} \label{subsec:HI gas Overdense Properties}
To investigate the \lya\ forest absorption lines, we do not use spectra in the wavelength range where damped \lya\ systems (DLAs) contaminate the spectra. 
Here, we search for DLAs in our spectra, performing DLA catalog matching and visual inspection.
As explained in \S\ \ref{subsec:Background Quasar Samples}, our spectra are taken from the three data sets of 
L13, BOSS DR12, and SDSS DR12. 
For the data set of L13, DLAs are already removed based on the DLA catalog of \cite{Noterdaeme2012a} who 
identify DLAs in the BOSS spectra by Voigt profile fitting.
For the data set of BOSS DR12, we find no DLAs in the \lya\ forest wavelength range in the \citeauthor{Noterdaeme2012a}'s DLA catalog.
For the data set of SDSS DR12, we perform visual inspection, and identify no DLAs.

Quasar host galaxies cause intrinsic strong metal absorption lines of {\sc SIV} $\lambda$1062.7, {\sc CIII} $\lambda$1175.7, {\sc NII} $\lambda$1084.0, and {\sc NI} $\lambda$1134.4 in the \lya\ forest wavelength range. We mask out the sufficient wavelength width $\pm$ 5\AA\ around these metal absorption lines.

The study of L13 has reproduced quasar continua by the MF-PCA continuum fitting technique \citep{Lee2012a}. 
The MF-PCA continuum fitting technique is essentially composed of two steps: (i) an initial PCA fit to the redward of the \lya\ emission line to reproduce the \lya\ forest continuum, and (ii) tuning the \lya\ forest continuum amplitude extrapolated to the blueward of the  \lya\  emission line with the cosmic \lya\ forest mean transmission of \cite{Faucher-Giguere2008a}.
We obtain quasar continua of the BOSS DR12 and the SDSS DR12 spectra by the MF-PCA technique 
with the code used in L13.
We include estimated median r.m.s, continuum fitting errors that are ($7$, $5.5$, $4.5$, $4\%$) for spectra with $\rm{S/N_{Ly\alpha}}=$ ($2-4$, $4-6$, $6-10$, $10 -15$) at $z \sim~ 2.5 - 3$ \citep{Lee2012a}. 
We calculate the \lya\ forest transmission $F(z)$ at each pixel:
\begin{eqnarray}
	F(z)= f(z)/C(z),
\end{eqnarray}
where $f(z)$ and $C(z)$ are the observed flux and the quasar continuum in the \lya\ forest wavelength range, respectively. 

We investigate the \lya\ forest absorption lines in the cylinders used by the galaxy overdensity calculation (\S\ \ref{subsec:Galaxy Overdensity}). We carry out binning for our spectra with the redshift range of $dz= 0.025(1+z)$ that corresponds to the length of the cylinder, and obtain an average \lya\ forest transmission $\langle F\rangle_{dz}$ and its error $\langle \sigma_F \rangle_{dz}$. 
Here, $\langle \sigma_F \rangle_{dz}$ is estimated with pixel noises in the spectra and continuum fitting errors.
The absorption of the  \lya\ forest is defined as $DA \equiv 1-\langle F\rangle_{dz}$.
We refer to the signal-to-noise ratio of the $DA$ detection S/N$_{\langle F\rangle}$ as $DA/\langle \sigma_F \rangle_{dz}$.
We calculate the S/N$_{\langle F\rangle}$ in the \lya\ forest wavelength range, and determine $z_h$ where $z_h$ is the redshift of the highest S/N$_{\langle F\rangle}$. 
We put the first cylinder centered at $z_h$ in each spectrum. We place additional cylinders that lie next to each other around the first cylinder.

To obtain statistically reliable results, we make use of the cylinders 
whose S/N$_{\langle F\rangle}$ is the highest in each sightline. However,
there is a possibility that this procedure for the cylinder placement
would bias the results. Here, we change the central redshift of the first cylinder 
from $z_h$ to the following two redshifts, I) and II). The two redshifts 
provide cylinders that cover I) the lowest and II) the highest wavelength ranges of the \lya\ forest.
Taking these different central redshifts of the cylinders, we find that 
our statistical results of Section \ref{sec:Galaxy-HI Correlation} 
change only by 4-5\%.

Each quasar sightline has 2-4 cylinders usable for our analysis. In total, there are 26 cylinders.
The number of cylinders is determined by the wavelength range where the following 2 ranges of i) and ii) overlap.
The two ranges are 
i) the \lya\ forest wavelength range, $1041(1+z{_{\rm{q}}}) {\rm{\AA}} \leq \lambda \leq 1185(1+z{_{\rm{q}}}) {\rm{\AA}}$ and 
ii) the $z{_{\rm{Ly\alpha}}}=2-3$ \lya\ forest absorption line range, $3647 {\rm{\AA}} \leq \lambda \leq 4862{\rm{\AA}}$, in the observed frame (\S\ \ref{subsec:Background Quasar Samples}).
For example, a quasar at $z{_{\rm{q}}}=3.36$ has $4539 {\rm{\AA}} \leq \lambda \leq 5166 {\rm{\AA}}$ for i). 
The wavelength range of the i) and ii) overlap is $4539 {\rm{\AA}} \leq \lambda \leq 4862 {\rm{\AA}}$. 
This wavelength range corresponds to $2.73 \leq z{_{\rm{Ly\alpha}}} \leq 3.0$. 
Because the length of a cylinder is $0.025(1+z)\sim0.1$ at this redshift range,
we obtain 2 cylinders from this sightline of the $z{_{\rm{q}}}=3.36$ quasar.

We use the data of the cylinders with S/N$_{\langle F\rangle} \geq 4$, and estimate both \delg\ and $\langle F\rangle_{dz}$  in the cylinders. 
Here we test whether this cut of S/N$_{\langle F\rangle} \geq 4$ gives impacts on our results.
We change the S/N$_{\langle F\rangle}$ cut from 4 to 3 and 5, 
and carry out the same analysis to evaluate how much different results 
can be obtained by the different S/N cuts.
We find that results of S/N$_{\langle F\rangle} \geq 3$ and $\geq 5$ cuts are very similar to those of S/N$_{\langle F\rangle} \geq 4$. Thus the different S/N cut within this range has a minimal impact on our results.
Note that the S/N$_{\langle F\rangle}$ cut below 3 raises the noise level and that the correlation signals are diminished.
Moreover, the S/N$_{\langle F\rangle}$ cut beyond 5 gives number of spectra too small to investigate the correlations.  

By these definitions and selections of the cylinders,
we have a total of 16 cylinders for the \delg\ - $\langle F\rangle_{dz}$ measurements.
In each cylinder, we calculate the \lya\ forest fluctuation whose negative values correspond to a strong \lya\ absorption:
\begin{eqnarray}
	\delta_{\langle F\rangle} \equiv \frac{\langle F\rangle_{dz}}{F_{\rm cos} (z)}-1,
\end{eqnarray}
where $F_{\rm cos} (z)$ is the cosmic \lya\ forest mean transmission. We adopt  $F_{\rm cos} (z)$ estimated by \cite{Faucher-Giguere2008a},
\begin{eqnarray}
	{F_{\rm cos} (z)}=\exp[-0.001845(1+z)^{3.924}].
	\label{eq:meanflux}
\end{eqnarray}
The error of \delF\ is estimated with the \lya\ forest transmission errors $\langle \sigma_F \rangle_{dz}$.

\begin{figure}[t]
	\centering
	\includegraphics[clip,bb=5 200 600 650,width=0.9\hsize]{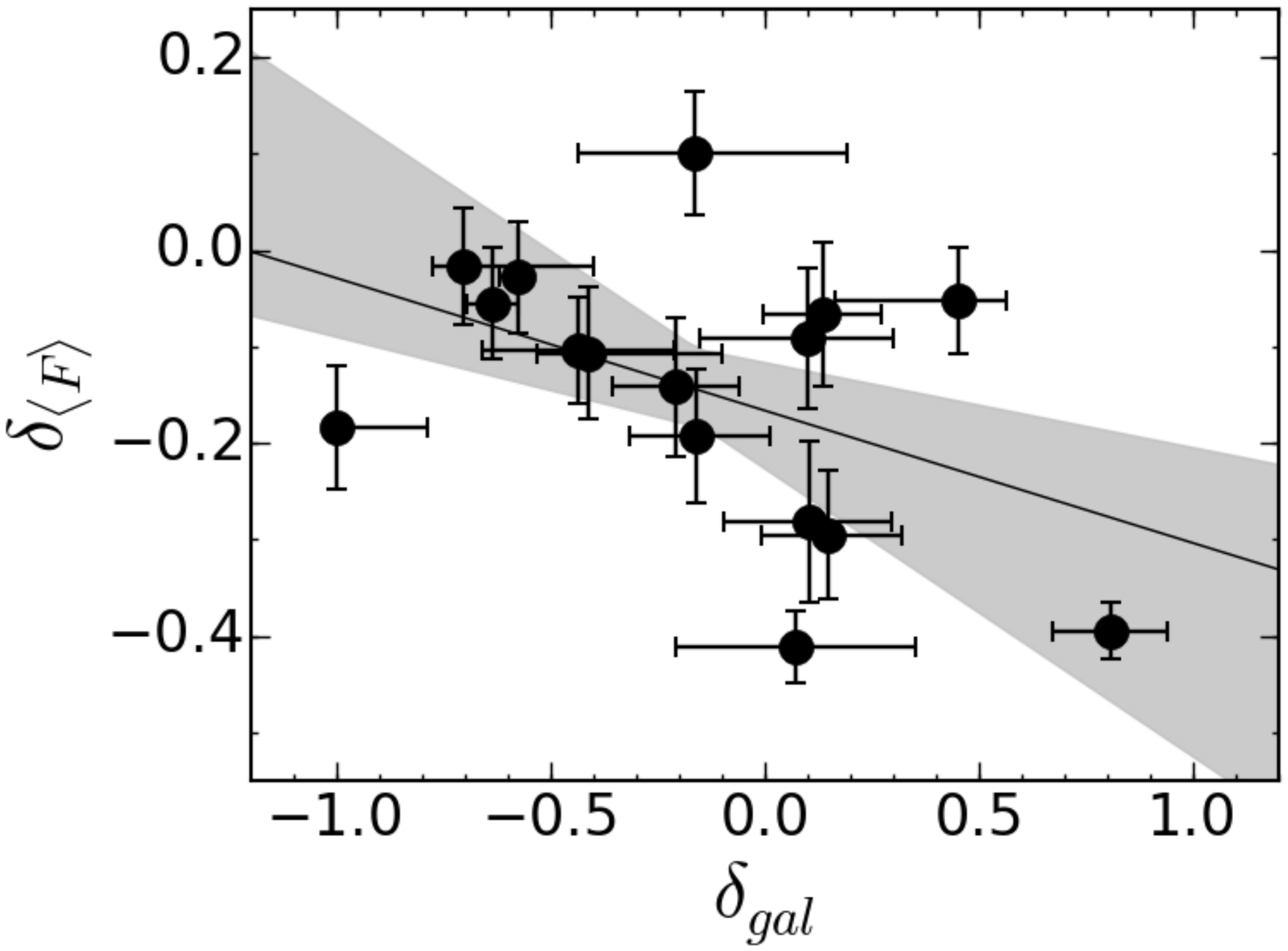}
	\caption{
	\lya\ forest fluctuation \delF\ as a function of the galaxy overdensity \delg. 
	The circles with the error bars represent the galaxy-\hi\ properties of the cylinders. 
	The best-fit linear model of Equation $\ref{eq:linear}$ is represented by a solid line,
	with the shaded region indicating the $1\sigma$ uncertainty range that is calculated by the perturbation method.
	}
	\label{fig:obs_results}
\end{figure}
\begin{figure}[t]
	\centering
	\includegraphics[clip,bb=5 200 600 650,width=0.9\hsize]{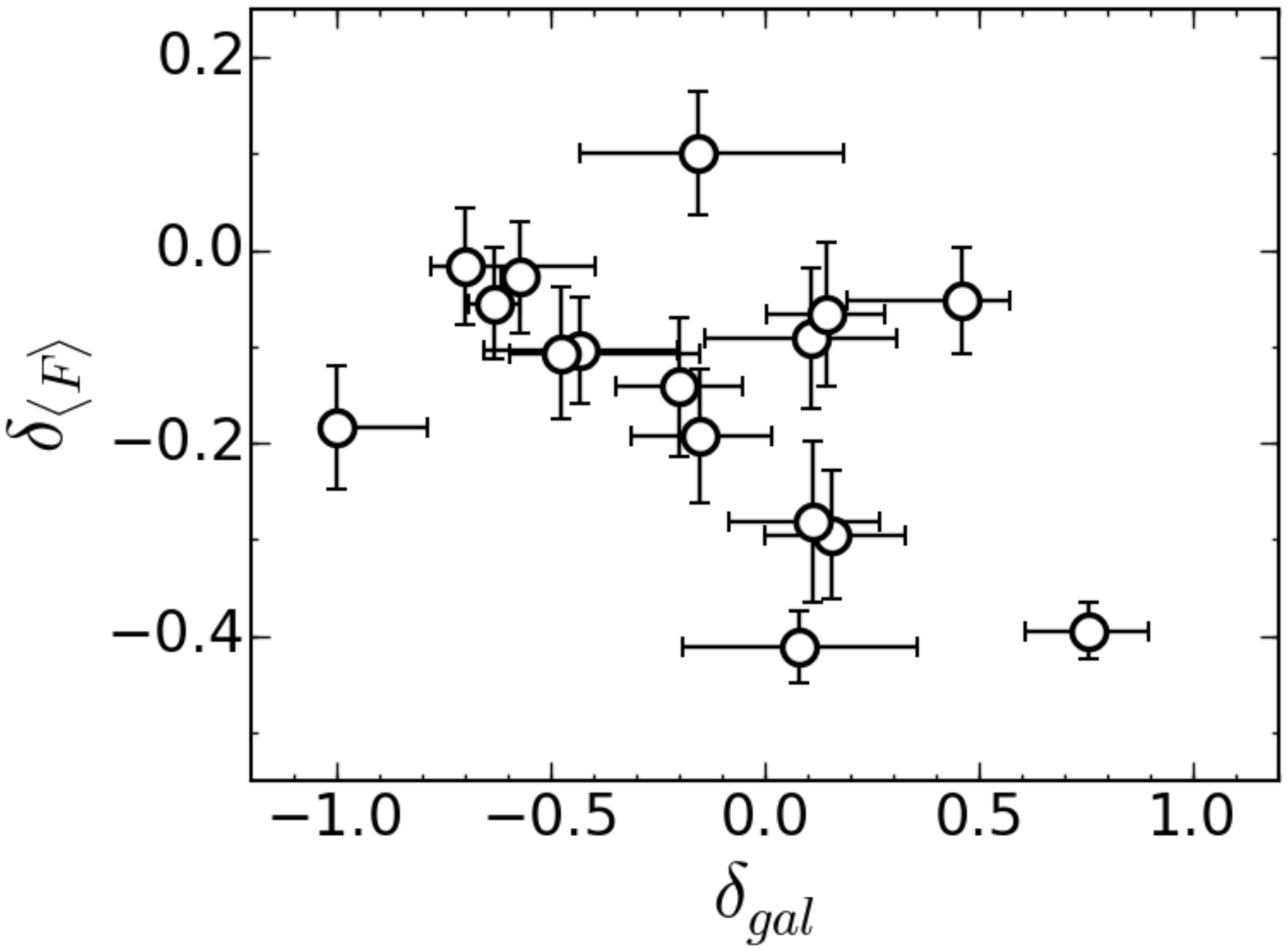}
	\caption{
	Same as Figure \ref{fig:obs_results}, but for revised \delg\ values with hollow cylinders. 
	The galaxy numbers of the revised \delg\ are estimated in hollow cylinders 
	whose inner and outer radii are $0.'4$ and $5'$, 
	which corresponds to 200 pkpc and 2.5 pMpc at $z\sim2.5$, respectively.
	}
	\label{fig:obs_annular}
\end{figure}

\section{Galaxy-IGM \hi\ Correlation} \label{sec:Galaxy-HI Correlation}
Figure $\ref{fig:obs_results}$ presents \delg\ and \delF\ values in the cylinders.
We calculate a Spearman's rank correlation coefficient $\rho_{\rm obs}$ of a nonparametric measure
to investigate the existence of a correlation between \delg\ and \delF. 
We obtain $\rho_{\rm obs} = -0.39$ that corresponds to the $\sim90\%$ confidence level
\footnote{
We find $\rho_{\rm obs} = -0.57$ that corresponds to the $\sim96\%$ confidence level, if we remove outliers (\S\ \ref{subsec:extreme value}).}
\citep{Wall2012a}.
We estimate the errors of $\rho_{\rm obs}$ by the perturbation method \citep{Curran2015a}.
We generate 1000 data sets of 16 cylinders whose data values include 
random perturbations following the gaussian distribution whose sigma is defined 
by the observational errors of the 16 cylinders.
We obtain 1000 Spearman's $\rho$ values for the 1000 data sets.
We define the error of $\rho_{\rm obs}$ as the range of 68 percentile distribution for the 1000 Spearman's $\rho$ values.
The estimated error of $\rho_{\rm obs}$ is 0.2 at the $1\sigma$ level. 
The $1\sigma$-error range of $\rho_{\rm obs}$ is thus $\rho_{\rm obs}=-0.6 - (-0.2)$. This range of $\rho_{\rm obs}$
corresponds to the confidence level range of $\sim 80-100$\%. In other words, 
the $\rho_{\rm obs}$ error changes the results of the $\sim90\%$ confidence level only by $\pm \sim10\%$.

We then apply chi-square fitting to the relation of \delg -\delF, and obtain the best-fit linear model,
\begin{eqnarray}
	\delta_{\langle F\rangle} = -0.17_{-0.06}^{+0.06} -0.14_{-0.16}^{+0.06} \times \delta_{\rm{gal}}
	\label{eq:linear}
\end{eqnarray}
that is shown with the solid line in Figure $\ref{fig:obs_results}$. 
Figure $\ref{fig:obs_results}$ and Equation ($\ref{eq:linear}$) suggest weak evidence of an anti-correlation between \delg\ and \delF.
The suggestive anti-correlation between \delg\ and \delF\ indicates that high-$z$ galaxies exist around an excess of \hi\ gas in the \lya\ forest. 

There is a possibility that the spatial correlation in the same sightlines would bias the results.
To evaluate this possible bias, we choose one cylinder for each sightline that does not have 
any spatial correlations with the other cylinders, and conduct the same analysis. 
We find an anti-correlation at the $\sim80\%$ confidence level that
falls in the $1\sigma$ error range of the results with all of the cylinders.
We thus conclude that the results do not change by the bias of the correlation 
that is not as large as the one of statistical uncertainties.

Strong \lya\ absorption lines can be made by the CGM of galaxies that lie near the quasar sightlines.
\cite{Rudie2012a} have studied velocities and spatial locations of \hi\ gas surrounding star-forming galaxies at $z\sim 2-3$, and found that 
the \hi\ column density rapidly increases with decreasing an impact parameter within 200 pkpc.
We investigate the \lya\ absorption lines associated with the CGM. 
In each cylinder used in \S\ \ref{subsec:HI gas Overdense Properties}, we calculate 
revised \delg\ values whose galaxy numbers are estimated in hollow cylinders whose inner radius is $0.'4$ corresponding to an impact parameter of 200 pkpc. 
We find that there are only $0-1$ galaxies in a $0.'4$ radius cylinder.
The white circles in Figure $\ref{fig:obs_annular}$ represent the hollow cylinder results that are very similar to the black circles in Figure $\ref{fig:obs_results}$.
Figure $\ref{fig:obs_annular}$ indicates no significant differences in the \delg-\delF\ distributions and the $\rho_{\rm obs}$ value, and suggests that the CGM of galaxies is not the major source of the small \delF\ values.

\section{SIMULATIONS} \label{sec:SIMULATIONS}
We perform cosmological hydrodynamical simulations with the RAMSES code \citep{Teyssier2002a} to investigate the spatial correlations of galaxies and IGM \hi\ of our observational results (Section \ref{sec:Galaxy-HI Correlation}). 
The initial conditions are generated with the COSMIC package \citep{Bertschinger1995a}, and are evolved using Zel'dovich approximation.
We include both dark matter and baryon using $N$-body plus Eulerian hydrodynamics on a uniform grid.  
The simulations are performed in a box size of $80h^{-1}$ cMpc length with $512^3$ cells and a spatial resolution of $156h^{-1}$ ckpc. We use $512^3$ dark matter particles with a mass resolution of $3.16 \times 10^8 M_{\odot}$. The mean gas mass per cell is  $5.4 \times 10^7 M_{\odot}$. 

We include the ultraviolet background model of \cite{Haardt1996a} at the reionization redshift $z_{\rm{reion}}= 8.5$. 
We investigate the gas temperature value at the mean gas density in the simulations, and find $T = 1.4\times10^4$K that is consistent with observational measurements at $z\sim 2-3$ \citep{Becker2011a}. 
We assume the photoionization equilibrium. We apply the optically thin limit, and do not produce any DLAs.
Note that our simulations do not include feedback effects on the \lya\ forest. Because the feedback mostly affects 
high-density absorbers with $N_{\rm HI} > 10^{16}$ cm$^{-2}$ \citep{Theuns2001a}, 
the lack of the feedback effects does not significantly change the large-scale correlation of 
galaxy overdensities and \lya\ forest absorption lines.

Dark matter haloes in the simulations are identified by the HOP algorithm \citep{Eisenstein1998a}. 
We use dark matter haloes containing more than 1000 dark matter particles.
We have compared the halo mass functions at $z\sim 2.5$ in our simulations with the halo mass function of the high resolution {\it N}-body simulations \citep{Reed2007a}, and found a good agreement 
within $\sim 30 \%$ in abundance. Our simulations resolve dark matter haloes with a mass of $\log_{10} M_h / M_{\odot} > 11$.

\subsection{Mock Galaxy Catalog} \label{subsec:Mock Galaxy Catalog}
We create mock galaxy catalogs from the simulations using the abundance matching technique 
\citep[e.g.,][]{Peacock2000a, Vale2004a, Moster2010a, Behroozi2013a}
that explains observational results of stellar mass functions.
We make simulated galaxies, populating each halo with one galaxy. We assume the stellar-to-halo mass ratio (SHMR) with a functional form
\begin{eqnarray}
	M_\ast/M_h=f_0\frac{(M_h/M_1)^\alpha}{1+(M_h/M_1)^{\alpha-\beta}},
	\label{eq:SHMR}
\end{eqnarray}
where $f_0$, $M_1$, $\alpha$, and $\beta$ are free parameters. We produce SMFs at $z\sim 2.5$ with many sets of these parameters. We compare these SMFs with observed SMFs \citep{Muzzin2013b, Tomczak2014a}, and find the best-fit parameter set reproducing the observed SMFs. The best-fit parameter set is ($M_1, \alpha, \beta, f_0) = (1.2\times10^{12} M_\odot, 1.0, -0.3, 0.04)$.
The SHMR with these best-fit parameters is consistent with the one estimated by \citet{Behroozi2013a} within the $1\sigma$ error levels.
We use the simulated galaxies whose stellar mass is $\log_{10} (M_* / M_{\odot}) > 9.7$ that is the same stellar mass limit in the COSMOS/UltraVISTA catalog (\S \ref{subsec:Photometric Galaxy Samples}). The stellar mass limit corresponds to the minimum halo mass of $M_h > 4.4 \times 10^{11} M_{\odot}$. The simulated galaxies consist of 2221 galaxies at $z=2.4-2.5$ that agree with the number of galaxies at $z=2.4-2.5$ in our COSMOS/UltraVISTA photometric samples.

\subsection{\lya\ Forest Catalog} \label{subsec:lya Forest Catalog}
In the simulation boxes, we make mock spectra along the random sightlines parallel to a principal axis defined as the redshift direction. 
The \lya\ transmitted flux is computed with the fluctuating Gunn-Peterson approximation \citep[FGPA; e.g.,][]{Weinberg1998a, Weinberg2003a, Meiksin2009a, Becker2015a}, because the FGPA method is simple and fast in computing. 
We ignore the gas velocities and the effect of redshift-space distortion on the \lya\ forest, 
testing whether this method gives reliable results. 
The FGPA is a good approximation for absorbers with densities around and below the cosmic mean \citep{Rakic2012a}.

We choose two typical sightlines, and conduct full optical depth calculations with gas velocities and redshift-space distortions \citep[e.g.,][]{Meiksin2015a, Lukic2015a}. 
We then compare the results of these two sightlines with those given by our original method. We find that the difference is only $< 6$\% 
that is not as large as the one of statistical uncertainties.

The FGPA gives the \lya\ optical depth as
\begin{equation}
\tau=\frac{c\sigma_\alpha \bar{n}_{\rm H}(z)}{\nu_\alpha H(z)}x_{\rm HI}\Delta_b\propto\Delta_b^{2-0.72(\gamma-1)}\label{eq:FGPA},
\end{equation}
where $\sigma_\alpha$ is the \lya\ cross section, $\bar{n}_{\rm H}(z)$ is the cosmic mean density of hydrogen atoms at redshift $z$, $\nu_\alpha$ is the Ly$\alpha$ resonance frequency, $H(z)$ denotes the Hubble constant at redshift $z$, $x_{\rm HI}$ is the fraction of neutral hydrogen,  $\Delta_b$ is the baryonic density in units of the mean density, and $\gamma$ is the power-law slope of the temperature-density relation in the IGM \citep{Hui1997a}.  The $\gamma$ value is set as $\gamma=1.5$ that is consistent with observations of the Ly$\alpha$ forest transmissions \citep{Becker2011a, Boera2014a}. 
We investigate the fidelity of our simulated \lya\ forest model. We compare the one-dimensional power spectrum of the transmitted flux in our simulations with the SDSS and the BOSS measurements \citep{McDonald2006a, Palanque-Delabrouille2013a}, and we find a good agreement.
We scale the mean  \lya\ transmitted flux to the cosmic \lya\ forest mean transmission of \cite{Faucher-Giguere2008a} (Equation (\ref{eq:meanflux})).
This scaling method is widely used in the literature \citep[e.g.,][]{White2010a, Lukic2015a}. 
We rebin the simulated spectra, and produce the SDSS and the BOSS pixel width of $69 {\rm km\ s^{-1}}$. We add Gaussian noises to the simulated spectra, accomplishing the $\rm{S/N_{Ly\alpha}}\sim5$ that corresponds to the typical $\rm{S/N_{Ly\alpha}}$ in our quasar samples  (\S \ref{subsec:Background Quasar Samples}).

\begin{figure}[]
	\centering
	\includegraphics[clip,bb=5 200 600 650,width=1.0\hsize]{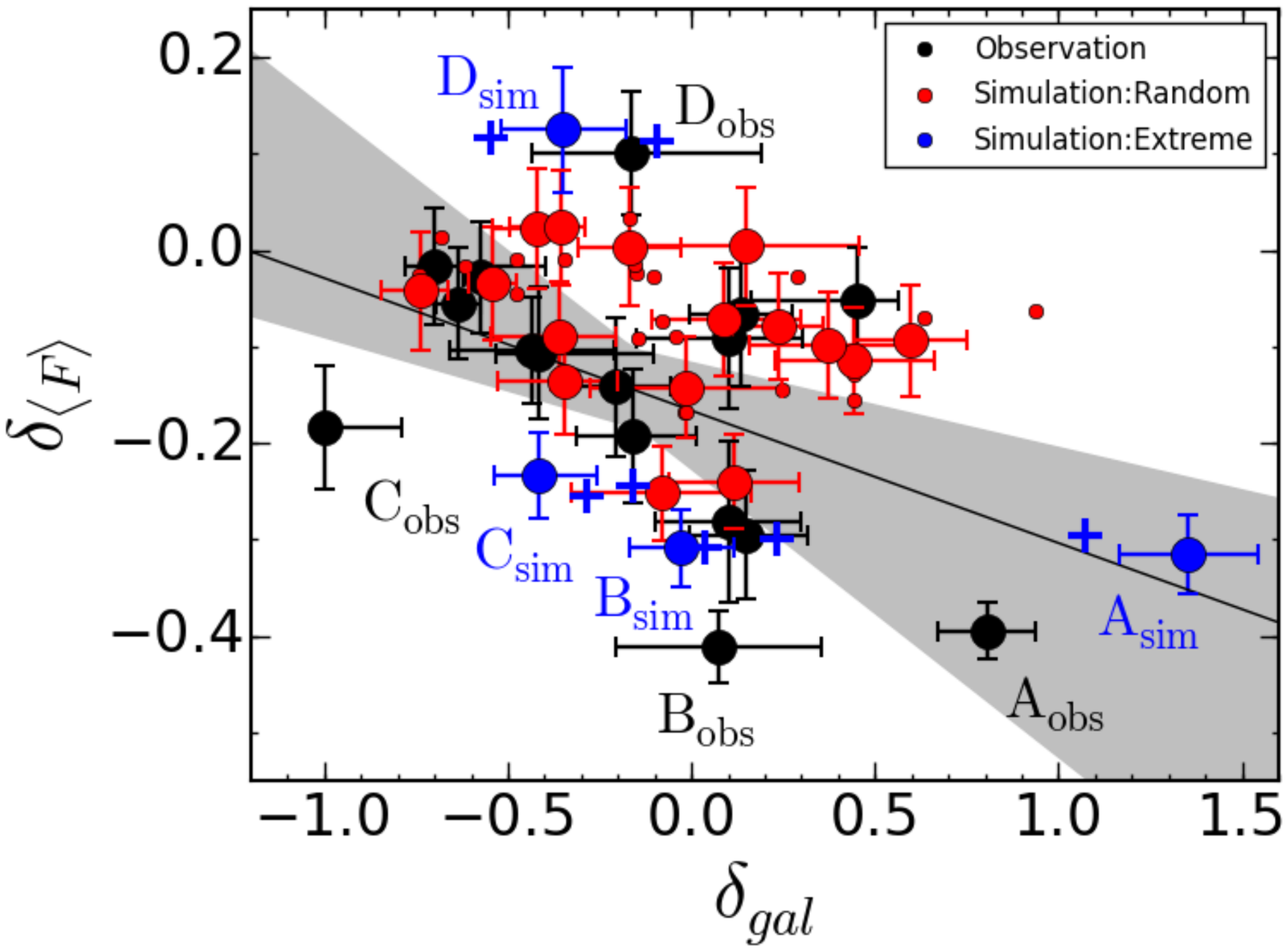}
	\caption{
	Same as Figure \ref{fig:obs_results}, but we overplot our simulation results.
	The black circles represent the observation data. The red dots are the simulation results. 
	The red circles show an example of 16 cylinders that are randomly chosen from our simulation results (red dots). 
	The blue circles present the simulated results for the four cylinders that have extreme values of \delg\ and/or \delF\ 
	similar to \Aobs-\Dobs\ (\S\ \ref{subsec:extreme value}). 
	The blue crosses are other counterparts found in our simulations (APPENDIX \ref{sec:APPENDIX C}).}
	\label{fig:overplot}
\end{figure}

\subsection{Simulated Galaxy-IGM \hi\ Correlation} \label{subsec:Simulated Galaxy-HI Correlation}
In Figure $\ref{fig:overplot}$, the red dots represent a cylinder of \delg\ and \delF\ in our simulations.  
We make 1000 sets of the 16 cylinders selected from the simulations, mocking our observed 16 cylinders.
The red circles in Figure \ref{fig:overplot} denote one example set of the 16 mock cylinders.
We fit a linear model to each of these 1000 sets of the 16 mock cylinders, and 
obtain Spearman's rank correlation coefficient $\rho_{\rm sim}$ values for the 1000 sets.
We find that a 1 $\sigma$ distribution of $\rho_{\rm sim}$ corresponds to the range of $-0.35-(-0.60)$
that indicates the existence of the anti-correlation between \delg\ and \delF in the simulations.
This $\rho_{\rm sim}$ distribution corresponds to the $\sim90\% \pm10\% $ confidence level
that is the same as our conclusions for the observational data (Section \ref{sec:Galaxy-HI Correlation}).

To test the convergence of our simulations, 
we perform simulations that have a box size of 80 $h^{-1}$ cMpc length with $256^3$ cells. 
We detail these results in APPENDIX \ref{sec:APPENDIX A}.

We test whether the noise distribution makes significant changes from our conclusions. We use the $\rm{S/N_{Ly\alpha}}$ probability distribution same as the one of our observational spectra of the 16 cylinders. We add noise to our 16 simulation spectra, following the $\rm{S/N_{Ly\alpha}}$ probability distribution. 
We calculate Spearman's rank correlation coefficient $\rho_{\rm sim}$ values for 16 simulation spectra with the noise.
Conducting this test for $\sim 10$ times, we find that the $\rho_{\rm sim}$ values are not different from our original result beyond the statistical errors.
 
We estimate how the correlation changes when the full redshift range of the observations is considered. 
From $z=2.5\ (3)$ to $z=2\ (2.5)$, the structure growth increases only by $\sim14\%$. The galaxy clustering and \lya\ forest 
clustering are expected to grow accordingly by $\sim10\%$. 
In Figure $\ref{fig:overplot}$, this redshift evolution shifts \delg\ (\delF) values of the simulation data points rightward (downward) 
only by $\sim10\%$ that is not as large as the one of the statistical uncertainties.

\section{DISCUSSION} \label{sec:DISCUSSION}
\subsection{Comparison between the Observation and the Simulation Results} \label{subsec:Comparison between the Observation and the Simulation Results}
Sections \ref{sec:Galaxy-HI Correlation} and \ref{subsec:Simulated Galaxy-HI Correlation} present the observation and the simulation results. Both observation and simulation results indicate weak evidence of the anti-correlation between \delg\ and \delF\ (Figure \ref{fig:overplot}).
Moreover, the Spearman's rank correlation coefficient $\rho_{\rm obs} =  -0.39$ from the observations (Section \ref{sec:Galaxy-HI Correlation}) falls in the $\rho_{\rm sim}$ range of $-0.35-(-0.60)$, indicating that simulations well reproduce observational results.

\begin{figure}[]
	\centering
	\includegraphics[clip,bb=5 80 600 800,width=1.0\hsize]{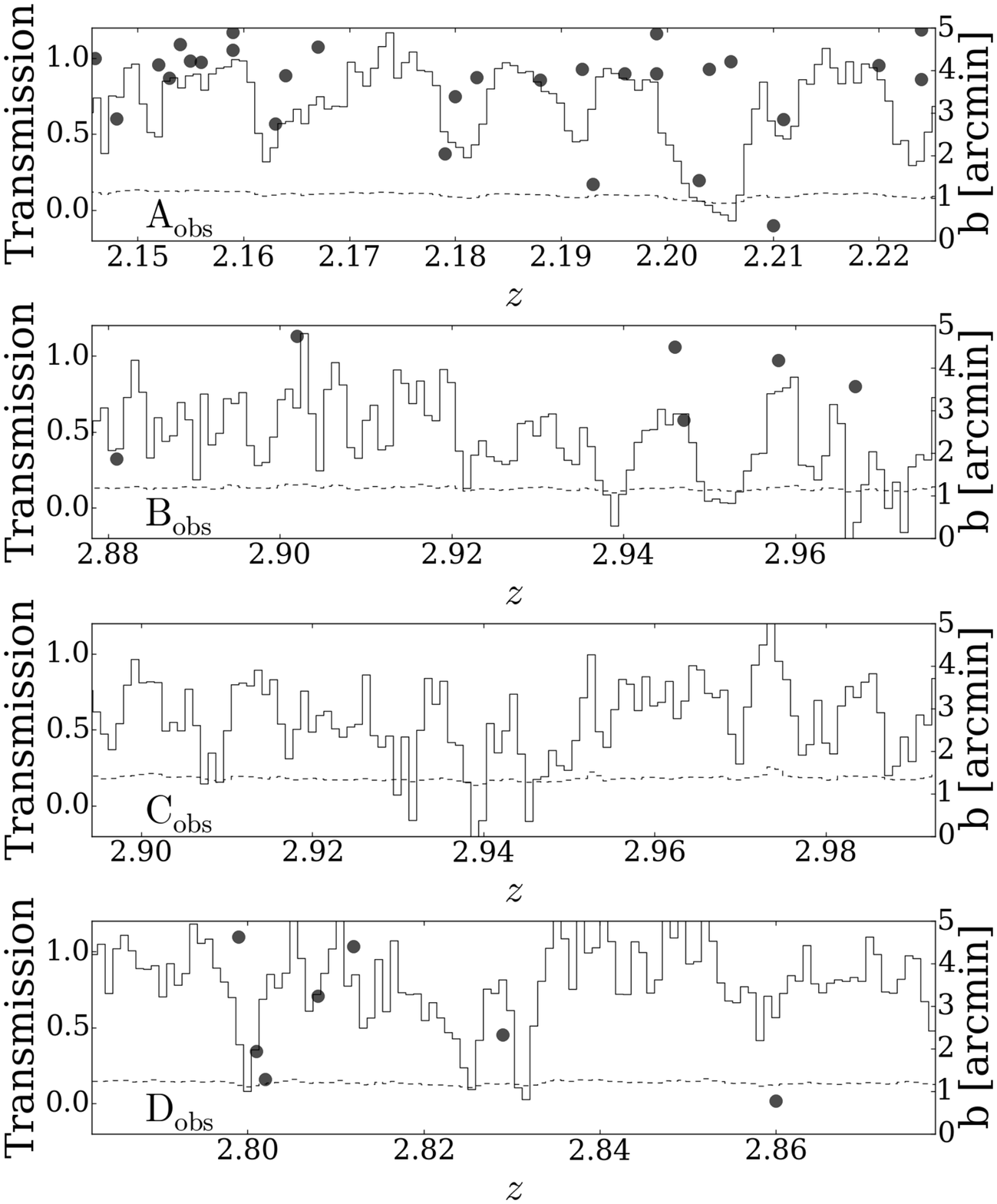}
	\caption{
	Observed background quasar spectra of the four cylinders that have extreme values of \delg\ and/or \delF\ 
	(\S\ \ref{subsec:extreme value}).
	The solid lines depict transmission per pixel as a function of redshift for the \lya\ forest absorption lines 
	(left-hand axis). 
	The dashed lines represent the noise per pixel.  
	The black points denote redshifts and impact parameters (right-hand axis) of our photo-$z$ galaxies in the cylinders.
	Note that the photometric redshift uncertainty is comparable to the full redshift range of the cylinder.
	}
	\label{fig:extreme_obs}
\end{figure}
\begin{figure*}[t]
	\centering
	\includegraphics[clip,bb=20 0 750 550,width=0.9\hsize]{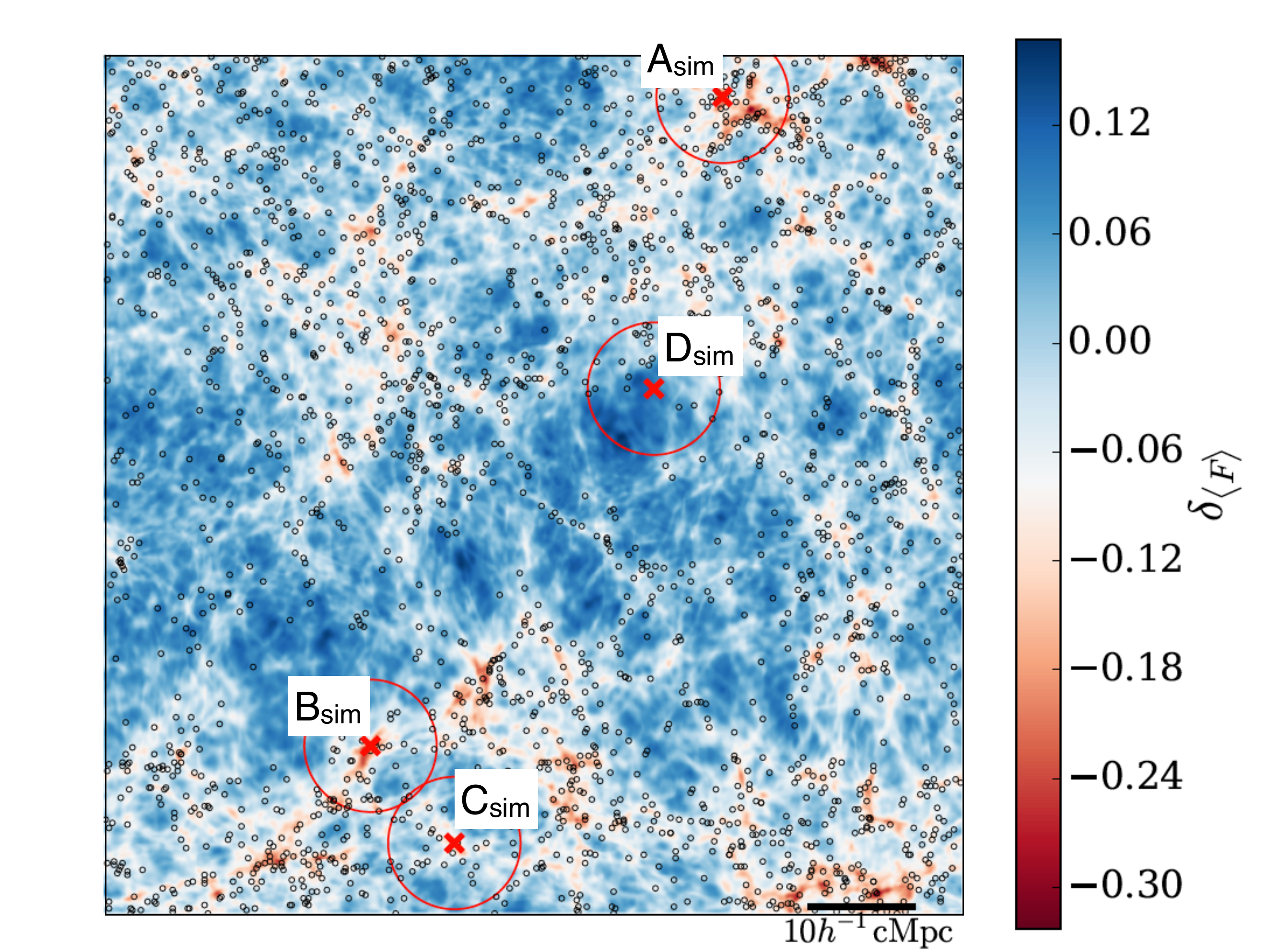}
	 \caption{
	 Projected sky map of the distribution of the mock galaxies and the \lya\ forest absorption lines.
	 The black circles represents mock galaxies whose stellar masses fall in $\log_{10} M_* / M_{\odot} > 9.7$. 
	 Note that the background color scale represents \delF\ defined in the full redshift range. 
	 The positions of sightlines are indicated by the red crosses enclosed by the red circles corresponding to the circumference of cylinders.
	  In the simulations, $10 h^{-1}$cMpc corresponds to $8.'5$ at $z\sim2.5$. 
	 }
   	 \label{fig:simulation skymap}
\end{figure*}
\begin{figure*}[t]
	\centering
	\includegraphics[clip,bb=0 10 500 320,width=0.8\hsize]{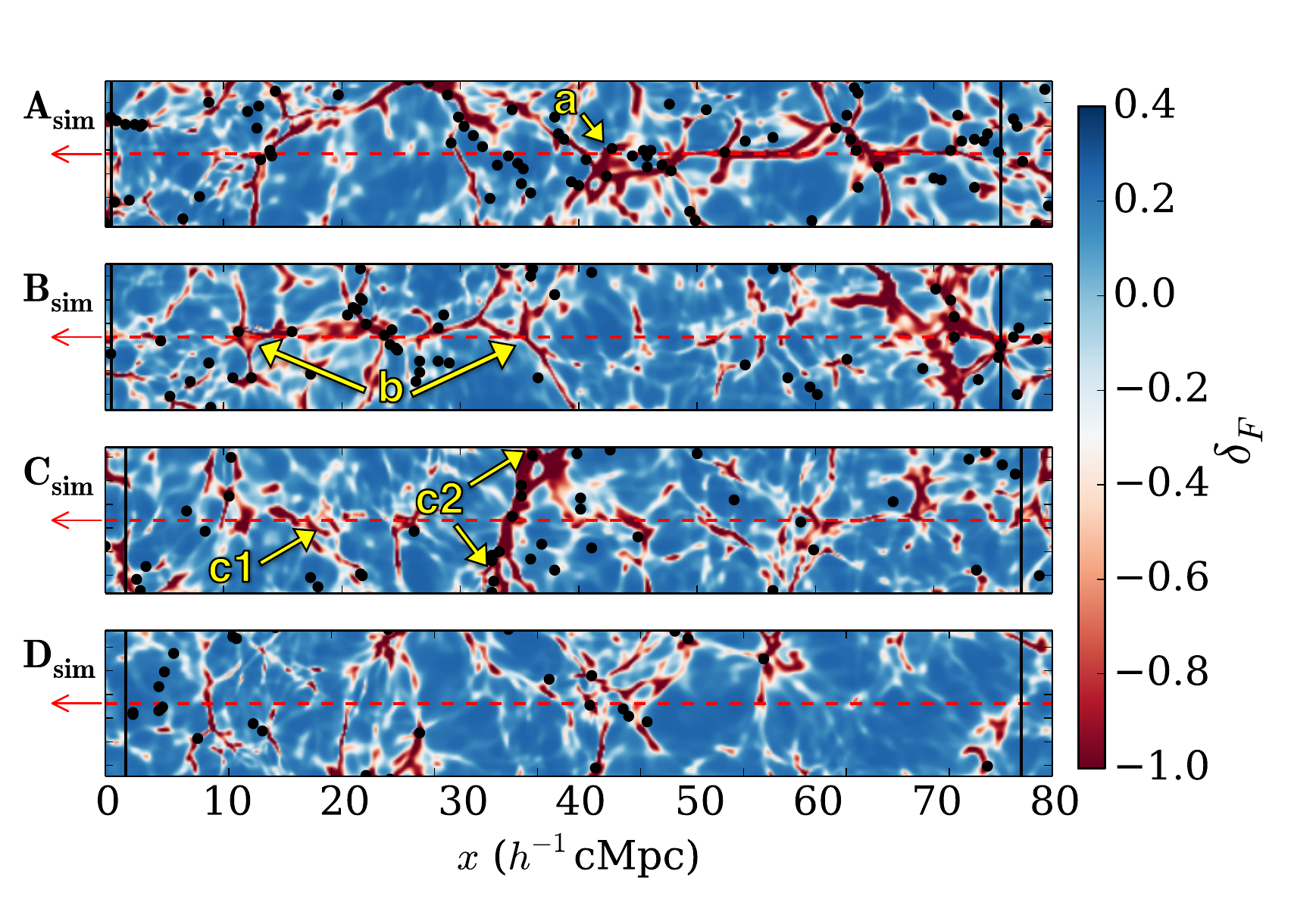}
	\caption{
	Maps of \delF\ around the sightlines of \Asim, \Bsim, \Csim, and \Dsim\ (\S\ \ref{subsec:extreme value}).
	The black points are projected positions of the mock galaxies in the cylinders.
	The background color scale represents a \delF\ value per pixel. 
	The black vertical lines represent the lower and upper edges of each cylinder.
	The red dashed lines represent the sightlines of \Asim, \Bsim, \Csim, and \Dsim. 
	The arrows indicate a galaxy overdensity in  \Asim (label 'a'),  two edges of gas filaments in \Bsim\ and \Csim\ (labels 'b' and 'c2'), and a galaxy void with gas filaments in \Csim 
	(label 'c1').
	}
	\label{fig:simulation sightline}  
\end{figure*}	
\begin{figure}[t]
	\centering
	\includegraphics[clip,bb=5 80 600 800,width=1.0\hsize]{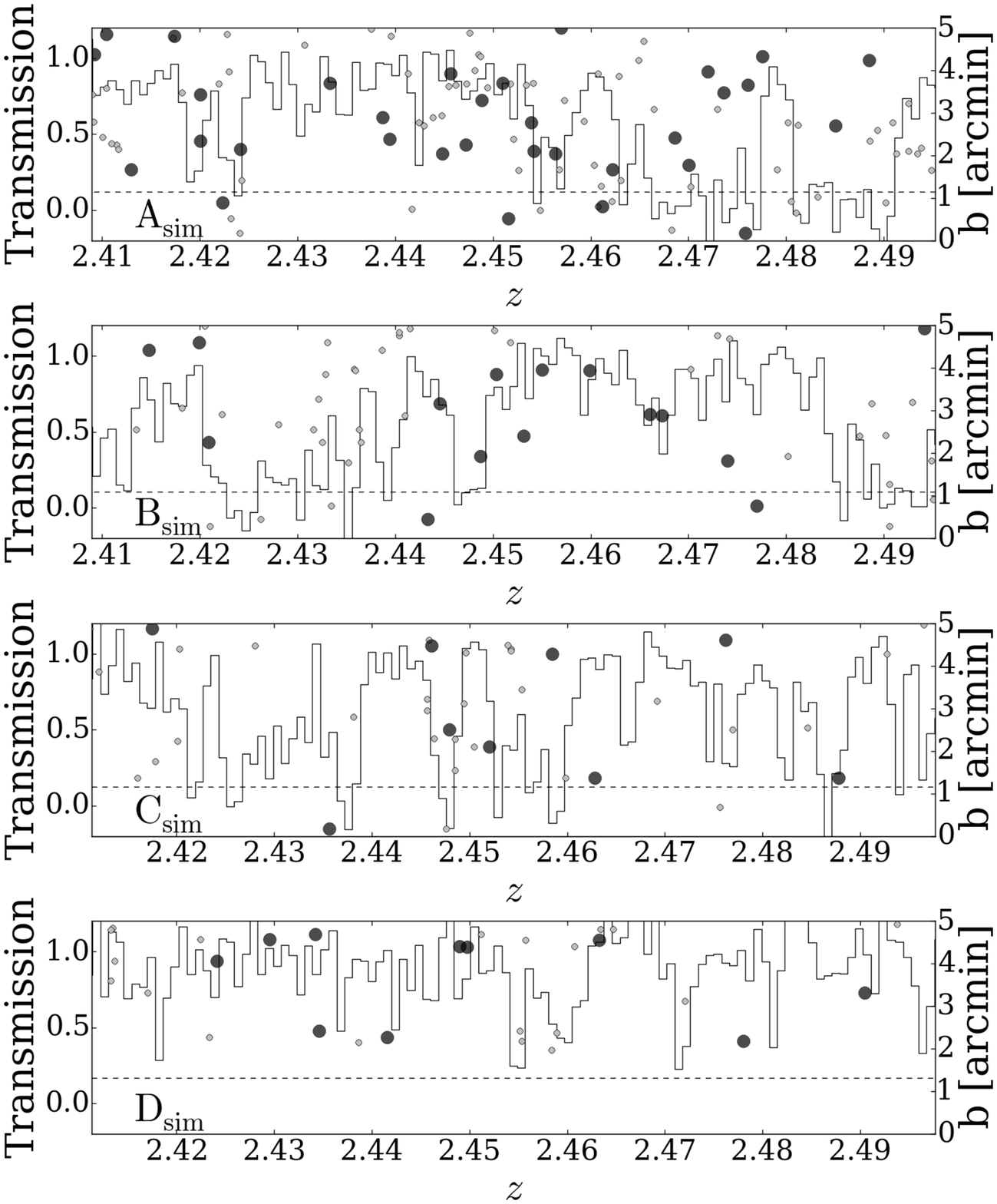}
	\caption{
	Same as Figure \ref{fig:extreme_obs}, but for the background quasar spectra in the simulations.
	Each panel corresponds to a cylinder that has extreme values of \delg\ and/or \delF\  
	in the simulations (\S\ \ref{subsec:extreme value}).
	The black points denote redshifts and impact parameters (right-hand axis) of our photo-$z$ galaxies in the cylinders.
	The gray points present photo-$z$ galaxies without photometric redshift uncertainties.
	}
	\label{fig:extreme_sim}  
\end{figure}

\subsection{Four Cylinders with an extreme value}\label{subsec:extreme value}
In Figure $\ref{fig:overplot}$, we find four cylinders with the labels of \Aobs, \Bobs, \Cobs, and \Dobs\ that have the largest (smallest) values of \delg\ or \delF\ among the observational data points.
Note that, in Figure $\ref{fig:overplot}$, \Aobs\ falls on the best-fit linear model within the errors, and that 
the other three cylinders, \Bobs, \Cobs, and \Dobs, show significant departures from the best-fit linear model by $0.1-0.2$ in 
$\delta_{\langle F\rangle}$ at the $2-3 \sigma$ significance levels.
These three cylinders weaken the anti-correlation signal found in Section \ref{sec:Galaxy-HI Correlation}. 
If we exclude these three cylinders, we obtain $\rho_{\rm obs} = -0.57$ that corresponds to the $\sim96\%$ confidence 
level.
\footnote{
In our simulations, there exist sets of cylinders that show significant ($>95\%$) confidence levels after we remove extreme cylinders, which agrees with our observational results.}

\par We present background quasar spectra of the four cylinders in Figure $\ref{fig:extreme_obs}$. The full width of the abscissa axis in the four panels of Figure $\ref{fig:extreme_obs}$ corresponds to the full redshift range of the cylinder. The black points in Figure $\ref{fig:extreme_obs}$ present the positions of galaxies with the best estimate of photometric redshifts and the impact parameter $b$ in reference to quasar sightlines.
We note that the photometric redshift uncertainty is comparable to the full redshift range of the cylinder.

We investigate the physical origin of the extreme \delg\ and/or \delF values in the four cylinders from our observations. We use the simulations performed in  Section 5, and search for cylinders in the simulations whose \delg -\delF\ values are similar to the four cylinders from our observations. With the simulation data, we create the sky map (Figure $\ref{fig:simulation skymap}$), and identify four mock cylinders that show \delg -\delF\ values most similar to \Aobs-\Dobs. We refer to these mock cylinders as the \Asim, \Bsim, \Csim, and \Dsim. The Maps of \delF\ around the sightlines of \Asim-\Dsim\ are shown in Figures $\ref{fig:simulation sightline}$ and $\ref{fig:extreme_sim}$, respectively. In Figure $\ref{fig:overplot}$, we overplot the \delg -\delF\ values of \Asim-\Dsim. Below, we describe properties and comparisons of \Aobs-\Dobs\  and \Asim-\Dsim.

{\bf{Cylinder A}}: 
\Aobs\  has the largest \delg\ and one of the smallest \delF\  values. The top panel of Figure $\ref{fig:extreme_obs}$ indicates that \Aobs\  has the largest number of galaxies and one of the strongest \lya\ forest absorption lines among the four cylinders of \Aobs-\Dobs. Interestingly, \Aobs\  coincides with one of the proto-cluster candidates reported by \cite{Chiang2014a}. The large \delg\ and the small \delF\ values of \Aobs\  suggest that a large galaxy overdensity is associated with the large amount of \hi\ gas \citep{Cucciati2014a, Chiang2015a, Cai2016a, Lee2016a}. Our simulation results in Figure $\ref{fig:simulation sightline}$ present that the sightline of \Asim\ penetrates gas filaments of LSSs and a galaxy overdensity like a proto-cluster at $z\sim2.46$ (label 'a').  Moreover, the top panel of Figure $\ref{fig:extreme_sim}$ indicates that \Asim\ would include a Coherently Strong \lya\ Absorption system (CoSLA) that traces massive overdensities on the scale of $\sim15 h^{-1}$cMpc \citep{Cai2016a}. 

Because there exists the high-quality spectra of \Aobs\ taken with VLT/X-shooter \citep{Vernet2011a}.
we show the characteristic of the \Aobs\ sightline based on the X-shooter spectrum 
in Appendix \ref{sec:APPENDIX B}.

{\bf{Cylinder B}}: 
\Bobs\  has the moderate \delg\ and the smallest \delF\ values. The second top panel of Figure $\ref{fig:extreme_obs}$ shows that \Bobs\  has the strong \lya\ forest absorption lines over the entire redshift range of the cylinder.  In our simulations, Figure $\ref{fig:simulation sightline}$ presents that the sightline of \Bsim\ goes through gas filaments at $z\sim2.43$ (label 'b'). 
Our simulations indicate that the sightline of \Bobs\  would penetrate gas filaments with the moderate number of galaxies.

{\bf{Cylinder C}}: 
\Cobs\  has the smallest \delg . Note that \Cobs\  has $\delta_{\rm{gal}}=-1$ corresponding to no galaxy in the cylinder. The second bottom panel of Figure $\ref{fig:extreme_obs}$ indicates that \Cobs\  does not have galaxies but the moderately strong \lya\ forest absorption lines. In our simulations, we identify \Csim\ that has \delg -\delF\  values most similar to those of \Cobs. Our simulation results in Figure $\ref{fig:simulation sightline}$ show that the sightline of \Csim\ penetrates a large void of LSSs at $z\sim2.43$ (label 'c1') and goes across gas filaments at $z\sim2.45$ (label 'c2'). Our simulations suggest that \Cobs\ would penetrate a large void, and go across gas filaments.
Note that \Csim\ has $\delta_{\rm{gal}}\sim-0.5$ that is larger than the \delg\ value of \Cobs. 
Our simulations find no cylinders with $\delta_{\rm{gal}}=-1$ and $ \delta_{\langle F\rangle}\sim-0.2$
that \Cobs\ has. 
This is the difference between \Cobs\  and \Csim. 
This difference is probably made, because
(1) photometric redshifts of galaxies in \Cobs\ have catastrophically large errors,
(2) there exist faint galaxies whose luminosities are just below the observational limit of $K_s = 23.4$ mag,
or  (3) a void of galaxies similar to \Cobs\ is missing in the limited box size of the simulations.

{\bf{Cylinder D}}: 
The cylinder with the  largest \delF\  is \Dobs.
The bottom panel of Figure $\ref{fig:extreme_obs}$ presents that \Dobs\  has the moderately weak \lya\ forest absorption lines.  In our simulations, Figure $\ref{fig:simulation sightline}$ shows that the sightline of \Dsim\  
crosses the low-density filaments. Our simulation results suggest that \Dobs would go through the orthogonal low-density filaments.

With the results of Figures \ref{fig:extreme_obs} and \ref{fig:extreme_sim}, we count the numbers of galaxies with photo-$z$ errors in each cylinder. We find 
(29, 6, 7) galaxies in (\Aobs, \Bobs, \Dobs), while there are (32, 15, 10) galaxies in (\Asim, \Bsim, \Dsim). These numbers agree within the $\sim 1-2 \sigma$ levels. 
However, the number of galaxies in \Cobs\ is 0 that is significantly smaller than the one of \Csim\ (see above for the difference of \Cobs\ and \Csim).

Note that there exist other counterparts of \Aobs-\Dobs\ in our simulations.
We detail these counterparts in APPENDIX \ref{sec:APPENDIX C}.

We investigate properties of other sightlines that do not have extreme values of \delg\ and \delF.
We find that these sightlines penetrate neither dense structures, filaments in parallel, nor large voids.

\subsection{Summary of the simulation comparisons} \label{subsec: Summary of the simulation comparisons}
In \S\ \ref{subsec:extreme value}, we discuss the physical origins of four cylinders (\Aobs, \Bobs, \Cobs, \Dobs ) that have extremely large (small) values of \delg\ and \delF. We use the simulations performed in Section \ref{sec:SIMULATIONS}, and identify four cylinders (\Asim, \Bsim, \Csim, \Dsim ) whose \delg -\delF\  values are close to (\Aobs, \Bobs, \Cobs, \Dobs ). The comparisons between \Aobs-\Dobs\  and \Asim-\Dsim\  suggest that sightlines in the observation would penetrate (1) a galaxy overdensity like a proto-cluster in \Aobs, (2) gas filaments in \Bobs, (3) a large void in \Cobs, and (4) orthogonal low-density filaments in \Dobs. 
In this way, our simulations provide the possible physical pictures of these four cylinders based on the structure formation models. 

The similarity between our observation and simulation results (Figure \ref{fig:overplot}) supports the standard picture of galaxy formation scenario 
in the filamentary LSSs \citep{Mo2010a} on which our simulations are based.

As noted in Section \ref{subsec:extreme value}, 
the three cylinders,  \Bobs, \Cobs, and \Dobs\
depart from the anti-correlation of \delg\ and \delF\ in Figure $\ref{fig:overplot}$.
Because the simulation counterparts of these three cylinders penetrate 
gas filaments, a large void, and orthogonal low-density filaments by chance,
the comparisons with our simulations suggest that the significant departures 
from the anti-correlation are produced by the filamentary LSSs and the observation sightlines. 
These chance alignment effects reduce the anti-correlation signal.

\section{SUMMARY} \label{sec:SUMMARY}
We investigate spatial correlations of galaxies and IGM \hi\ with the 13,415 photo-$z$ galaxies 
and the \lya\ forest absorption lines of the background quasars with no signature of damped \lya\ system contamination  in the 1.62 deg$^2$ COSMOS/UltraVISTA field. 
The results of our study are summarized below.

\begin{enumerate}
\item 
We estimate the \lya\ forest fluctuation \delF\ and
the galaxy overdensity \delg\ within the impact parameter of $2.5$ pMpc 
from the quasar sightlines at $z\sim 2-3$. 
We identify an indication of anti-correlation between \delg\ and \delF\ values (Figure \ref{fig:obs_results}). 
The Spearman's $\rho$ value of $-0.39$ indicates that there is weak evidence of an anti-correlation between 
\delg\ and \delF\ at a $\sim90\%$ confidence level. 
This anti-correlation suggests that high-$z$ galaxies are found in the excess regions of \hi\ gas in the \lya\ forest.

\item 
We perform cosmological hydrodynamical simulations with the RAMSES code, and identify an anti-correlation between \delg\ and \delF\ values in our simulation model that is similar to the one found in our observational data. We estimate the Spearman's $\rho$ for the \delg\ and \delF\ values in our simulation results that suggest the anti-correlation agreeing with the observational results.

\item
In our observational data, we identify four cosmic volumes that have very large or small values of \delg\ and \delF\ that are dubbed \Aobs, \Bobs, \Cobs, and \Dobs\  (Figure \ref{fig:overplot}). Three out of these four cylinders,  \Bobs, \Cobs, and \Dobs, present significant departures from the anti-correlation of \delg\  and  \delF, and weaken the signal of the anti-correlation.

\item 
In our simulations, we identify model counterparts of \Aobs, \Bobs, \Cobs, and \Dobs\ 
in the \delg\ and \delF\ plane (Figure  \ref{fig:overplot}), which
are referred to as \Asim, \Bsim, \Csim, and \Dsim, respectively.
The comparisons of \Aobs-\Dobs\ with \Asim-\Dsim\ indicate 
that the observations pinpoint (1) a galaxy overdensity like a proto-cluster in \Aobs, 
(2) gas filaments lying on the quasar sightline by chance in \Bobs, 
(3) a large void in $\rm{C_{obs}}$, 
and 
(4) orthogonal low-density filaments in \Dobs. 
Our simulations suggest that the three cylinders, \Bobs, \Cobs, and \Dobs\ significantly departing from the anti-correlation are produced by the filamentary LSSs and the observation sightlines. The chance alignment effects reduce the anti-correlation signal of \delg\ and \delF.
\end{enumerate}

The large-scale correlation of \delg-\delF\ found in Section \ref{sec:Galaxy-HI Correlation} is relatively weak. This is because the correlation is based on the relatively large cylinder whose redshift range is limited in photo-$z$ accuracy \citep[see Figure 1 of][]{Cai2016a}.
The small-scale galaxy-\hi\ relations can be studied with galaxies with spectroscopic redshifts. Here, the Hobby-Eberly Telescope Dark Energy Experiment (HETDEX) survey \citep{Hill2016a} will carry out the wide-field observations, and provide 0.8 million galaxies in $\sim 400$\ deg$^2$. The HETDEX survey will reveal the galaxy-\hi\ relations in the large cosmological volumes including a number of proto-cluster cadidates, gas filaments, and voids of LSSs. The galaxy-IGM relation study with 
HETDEX will be complementary to the programs of the MAMMOTH \citep{Cai2016a} and the CLAMATO \citep{Lee2014b, Lee2016a}.

\begin{acknowledgements}
We are grateful to Toru Misawa, Suzuka Koyamada, Toru Yamada, Ken Mawatari, Takashi Kojima, Andreas Schulze,   
Henry Joy McCracken, Rich Bielby, and Shun Arai
for their useful comments and constructive discussions. We are also grateful to Khee-Gan Lee for kindly providing the MF-PCA code.

Based on data products from observations made with ESO Telescopes at the La Silla Paranal Observatory under ESO programme ID 179.A-2005 and on data products produced by TERAPIX and the Cambridge Astronomy Survey Unit on behalf of the UltraVISTA consortium.

The COSMOS/UltraVISTA $K_s$-band selected galaxy catalog used in this work is compiled by \cite{Muzzin2013a}. 
The catalog contains PSF-matched photometry in 30 photometric bands covering the wavelength range 0.15$\micron$ $\rightarrow$ 24$\micron$ and includes the available $GALEX$ \citep{Martin2005a}, CFHT/Subaru \citep{Capak2007a}, UltraVISTA \citep{McCracken2012a}, S-COSMOS \citep{Sanders2007a}, and zCOSMOS \citep{Lilly2009a} datasets.

Funding for SDSS-III has been provided by the Alfred P. Sloan Foundation, the Participating Institutions, the National
Science Foundation, and the U.S. Department of Energy Office of Science. The SDSS-III Web site is http://www.sdss3.org/.
SDSS-III is managed by the Astrophysical Research Consortium for the Participating Institutions of the SDSS-III Collaboration
including the University of Arizona, the Brazilian Participation Group, Brookhaven National Laboratory, University
of Cambridge, Carnegie Mellon University, University of Florida, the French Participation Group, the German Participation
Group, Harvard University, the Instituto de Astrofisica de Canarias, the Michigan State/Notre Dame/JINA Participation
Group, Johns Hopkins University, Lawrence Berkeley National Laboratory, Max Planck Institute for Astrophysics, Max Planck
Institute for Extraterrestrial Physics, NewMexico State University, New York University, Ohio State University, Pennsylvania
State University, University of Portsmouth, Princeton University, the Spanish Participation Group, University of Tokyo, University
of Utah, Vanderbilt University, University of Virginia, University of Washington, and Yale University.

This work has made use of the services of the ESO Science Archive Facility.
Based on data products from observations made with ESO Telescopes at the La Silla Paranal Observatory under programme ID 086.A-0974. 

This work is supported by World Premier International Research Center Initiative (WPI Initiative), MEXT, Japan, and KAKENHI (15H02064) Grant-in-Aid for Scientific Research (A) through Japan Society for the Promotion of Science (JSPS).
\end{acknowledgements}

\appendix
\section{A. Convergence Test} \label{sec:APPENDIX A}
\begin{figure}[h]
	\centering
	\includegraphics[clip,bb=0 15 600 415,width=0.45\hsize]{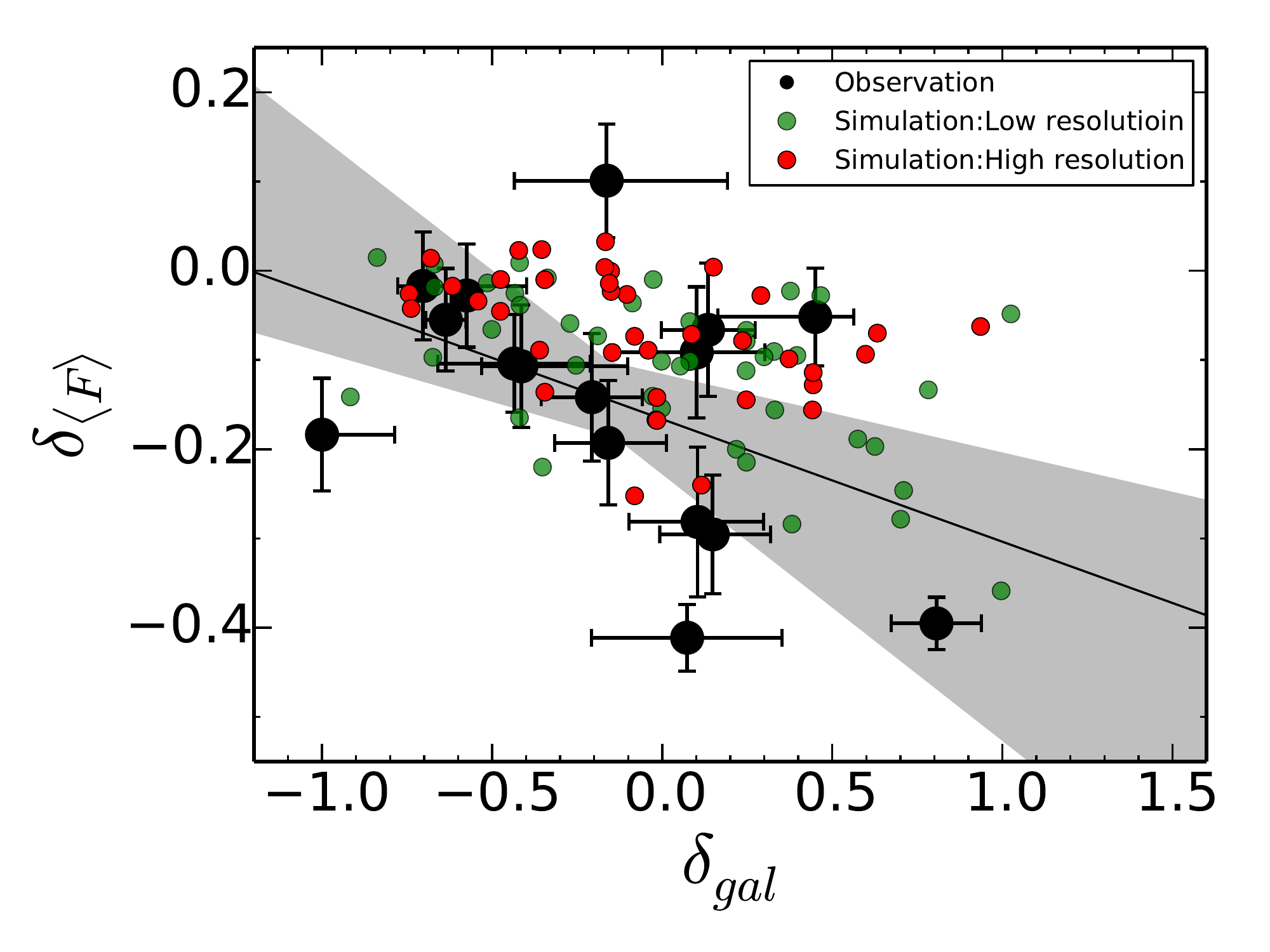}
	\caption{
	Same as Figure \ref{fig:obs_results}, but for the different resolution results with $256^3$ (green dots) and $512^3$ (red dots) cells.
	}
	\label{fig:conversion test}
\end{figure}

We perform simulations that have a box size of 80 $h^{-1}$ cMpc length with $256^3$ cells. 
Figure $\ref{fig:conversion test}$ presents \delg\ and \delF\ values in the cylinders.
The green (red) dots represent a cylinder of \delg\ and \delF\ in our simulations with $256^3$($512^3$) cells. 
In the same manner as our simulations with $512^3$ cells (\S\  \ref{subsec:Simulated Galaxy-HI Correlation}), 
we calculate a Spearman's rank correlation coefficient of the $256^3$-cell simulations $\rho_{\rm 256}$,
using 16 cylinders that are randomly chosen from the $256^3$-cell simulation results.
We obtain $\rho_{\rm 256} \sim -0.4$ that corresponds to the $\sim90\%$ confidence level. 
We find that results are very similar to those of our simulations with $512^3$ cells.

\section{B. Detail Properties of the A$_{\rm \lowercase{obs}}$ sightline}\label{sec:APPENDIX B}
In this Appendix, we show the supplementary VLT/X-shooter observations for the background quasar of \Aobs. 
The X-shooter observations were carried out in the service mode on 2010 December 31 (Program ID: 086.A-0974, PI: S. Lilly). 
We use the reduced X-shooter spectra 
that are publicly available on the European Southern Observatory (ESO) Science Archive Facility.\footnote{http://archive.eso.org} 
We select the spectra of the UVB and VIS arms, 
which cover the wavelength ranges of $3000$--$5595$\AA\ and $5595$--$10240$\AA, respectively. 
The spectral resolutions of the UVB and VIS arms are medium high, $R\sim4000$ and $R\sim7400$, respectively.
The observational details are summarized in Table \ref{tab:observations}. 

\begin{figure}[]
	\centering
	\includegraphics[clip,bb=80 0 1100 350,width=0.7\hsize]{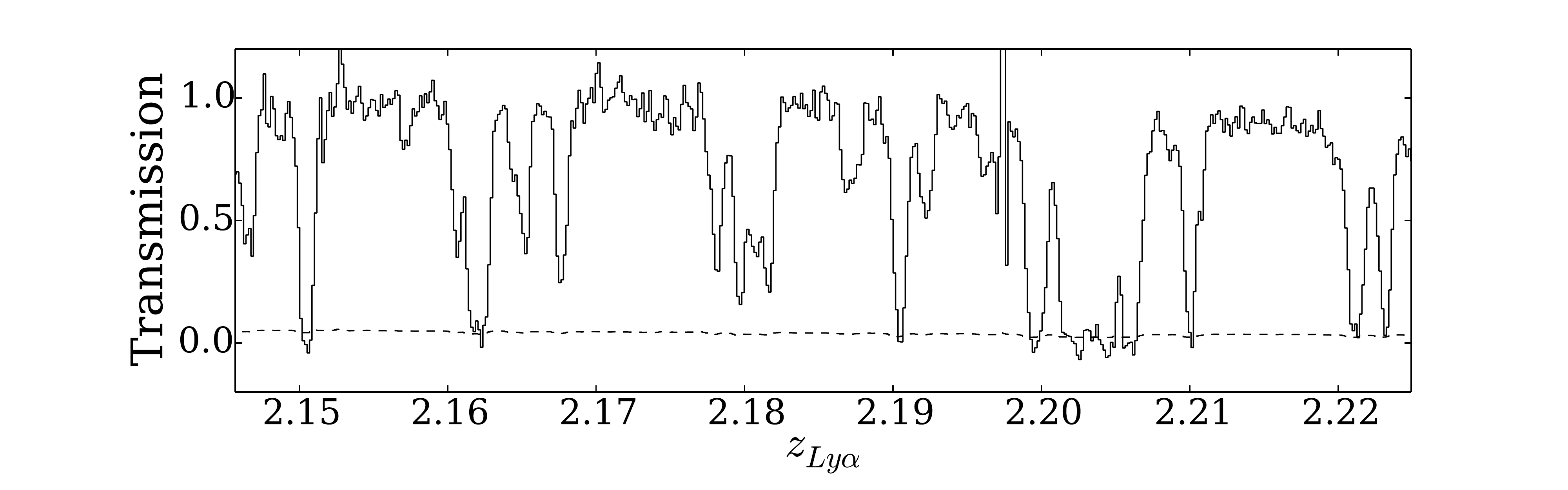}
	\caption{
	VLT/X-shooter spectrum of the background quasar of \Aobs.
	The solid lines represent transmission per pixel as a function of redshift for the \lya\ forest absorption lines. 
	The dashed curve denotes the noise per pixel.  
	}
  \label{fig:X-shooter}
\end{figure}

\begin{figure}[]
 \begin{minipage}{0.42\hsize}
	\centering
	\includegraphics[clip,bb=0 10 350 400,width=0.7\hsize]{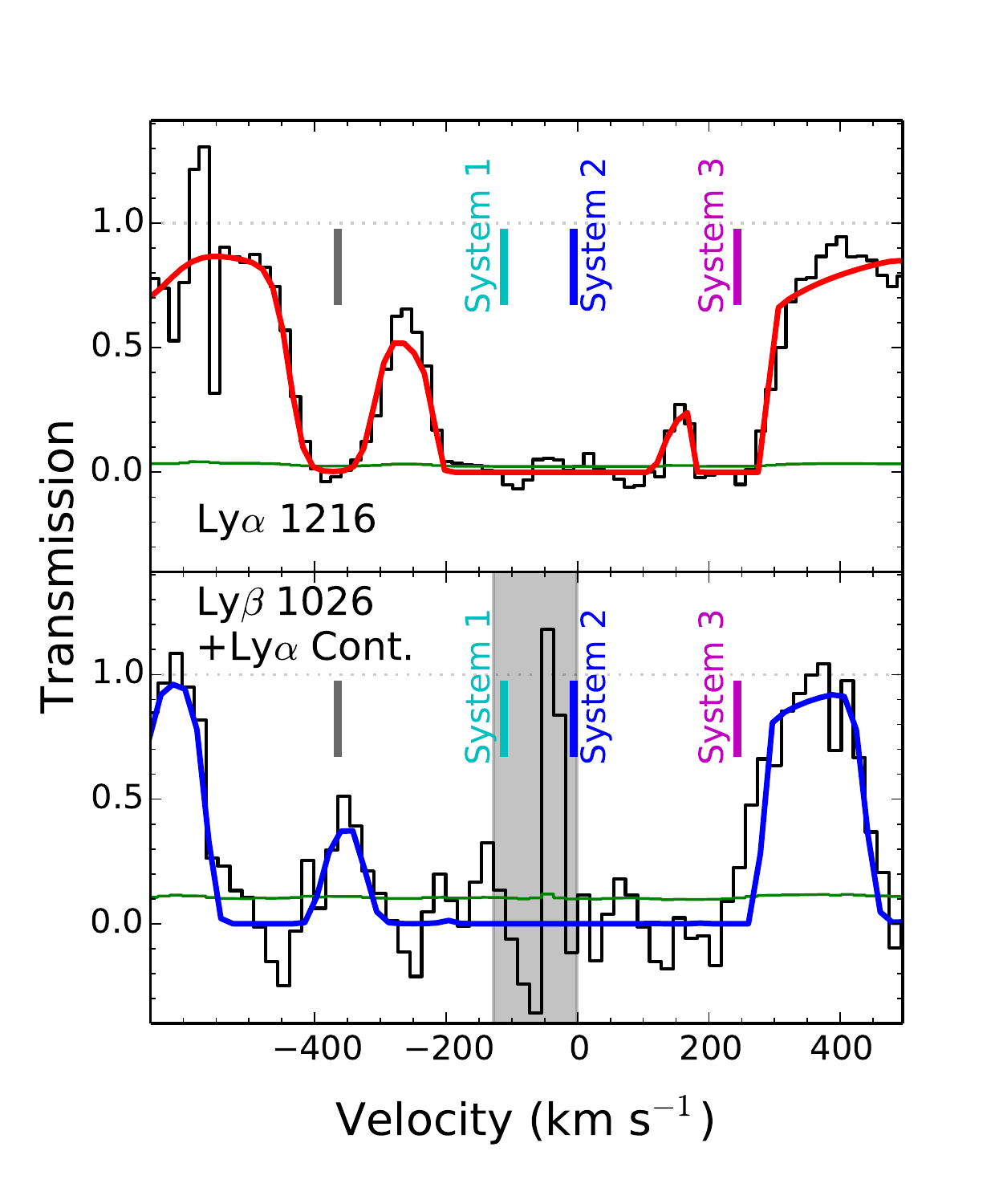}
 \end{minipage}
 \begin{minipage}{0.58\hsize}
	\centering
	\includegraphics[clip,bb=60 0 700 350,width=1.0\hsize]{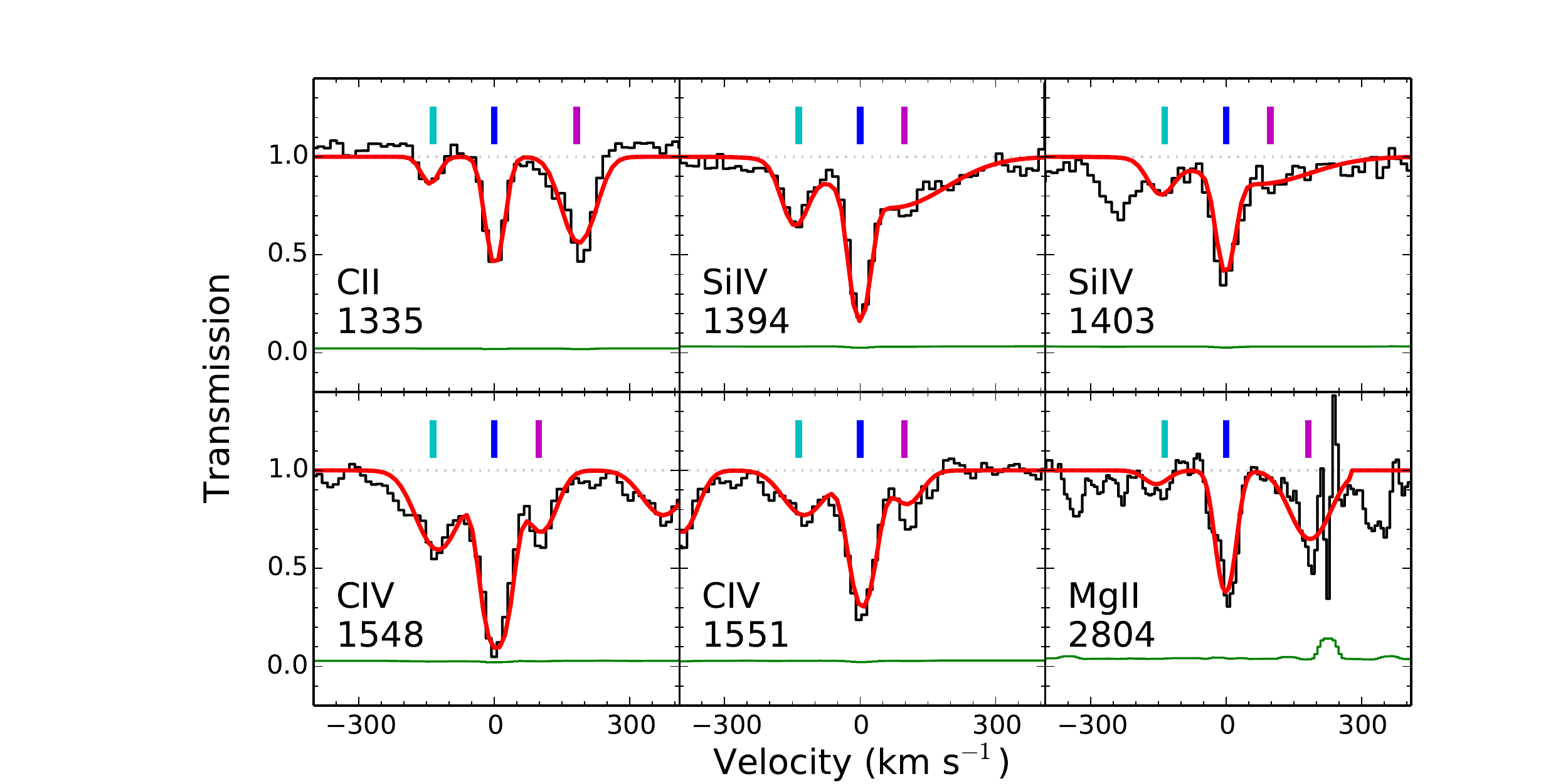}
 \end{minipage}
  \caption{
  		Left panel: 
		Zoom-in X-shooter spectrum around 
		the $z \sim 2.203$ {\sc Hi} \lya\ absorption systems (top) and \lyb\ absorption systems
		with $z \sim 1.70$ Ly$\alpha$ absorption contaminations (bottom).
		The black (green) histogram presents the transmission ($1\sigma$ noise) per pixel.
		The Voigt-profile decompositions (red curve) are made of the three \lya\ absorption systems, 
		System 1 (cyan ticks), System2 (blue ticks), System 3 (magenta ticks), and
		a \lya\ absorption system that has no metal lines (gray ticks). 
		The blue curve shows the fitting results for the  \lyb\ absorption systems 
		and $z \sim 1.70$ \lya\ absorption contaminations.
		Note that we simultaneously fit $z \sim 2.203$ \lya\ and \lyb\ absorbers 
		and $z \sim 1.70$ \lya\ absorption contaminations.
		The decompositions indicate that System 2 is the LLS at $z = 2.203$. 
		The gray shaded region in the bottom panels presents the wavelength range
		where we find a large sky subtraction systematic errors.
		Right panel:
		Same as the left panel, but for metal absorption lines associated with the \lya\ absorption systems
		 at $z \sim 2.203$.
		}
  \label{fig:X-shooterLLS}
\end{figure}

Figure \ref{fig:X-shooter} shows the X-shooter spectrum of the background quasar of \Aobs\ 
in the same wavelength range as Figure \ref{fig:extreme_obs}. 
Although the spectral resolution of the X-shooter spectrum is significantly higher than that of the BOSS spectrum (Figure \ref{fig:extreme_obs}),
we confirm that the sightline of \Aobs\ has \lya\ forest absorption lines with no signature of DLAs.
The only exception is the absorbers at $z \sim 2.203$, where saturated  \lya\ absorption lines are detected. 
The top left panel of Figure \ref{fig:X-shooterLLS} presents a zoom-in X-shooter spectrum around the saturated \lya\ absorption lines. 
Note that their \lyb\ absorption lines are also covered by the X-shooter spectrum. 
However, as shown in the bottom left panel of Figure \ref{fig:X-shooterLLS}, the \lyb\ absorption lines are also saturated. 
Moreover, the \lyb\ absorption lines are contaminated by foreground Ly$\alpha$ absorbers at $z \sim 1.70$,
which makes it difficult to precisely measure {\sc Hi} column densities of the {\sc Hi} absorption lines. 

As shown in the right panel of Figure \ref{fig:X-shooterLLS},
we identify metal absorption lines of {\sc Cii}, Si{\sc iv}, {\sc Civ} and Mg{\sc ii} at $z \sim 2.203$ 
in the X-shooter spectrum. 
There are three metal absorptions dubbed Systems 1, 2 and 3 that are labeled in Figure \ref{fig:X-shooterLLS}.
We fit Voigt profiles to these metal lines with {\sc vpfit}\footnote{http://www.ast.cam.ac.uk/\~{}rfc/vpfit.html} 
to measure column densities.  
The best-fit profiles for the metal lines are presented in the right panel of Figure \ref{fig:X-shooterLLS}.
Table \ref{tab:fitting} summarizes the measured column densities.  

To estimate {\sc Hi} column densities of Systems 1-3,
we first make photoionization models with the input observational measurements of 
{\sc Cii}, Si{\sc iv}, {\sc Civ}, and Mg{\sc ii} with those predicted from photoionization models. 
We perform multi-phase photoionization calculations with version $13.04$ of the {\sc cloudy} software \citep{Ferland2013a}. 
We conduct the {\sc cloudy} modeling for high-ionization phase clouds ({\sc Civ}, Si{\sc iv}) and low-ionization phase clouds (Mg{\sc ii}, {\sc Cii}) \citep[e.g.][]{Misawa2008a}.
We model these clouds in each phase, 
assuming a gas slab exposed by a uniform ultraviolet background \citep{Haardt2012a}
with a range of ionization parameters $U  \equiv \Phi/cn_{\rm H}$ ($-4.2 <  \log U < -0.6$), 
where $n_{\rm H}$ and $\Phi$ are the hydrogen volume density and  the ionizing photon flux incident on the gas cloud, respectively.
The solar relative abundances of \cite{Asplund2009a} are assumed.  
We then search for the best-fit model that minimizes $\chi^2$ 
between the measured metal column densities and the photoionization model predictions. 
We find that the best-fit models for Systems 1, 2, and 3 
have {\sc Hi} column densities of $\log N_\mathrm{H_I}$ (cm$^{-2}$) $\sim 16.0$, $19.0$, and $16.5$, respectively. 
We then fit Voigt profiles 
to the spectrum in the wavelength ranges of the \lya\ absorption lines (top left panel of Figure \ref{fig:X-shooterLLS}) and \lyb\ absorption lines (bottom left panel of Figure \ref{fig:X-shooterLLS}) 
by using the column densities of the {\sc cloudy} model results.
Because the spectrum in the \lyb\ wavelength range is contaminated by the foreground Ly$\alpha$ absorbers at $z \sim 1.70$,
we conduct simultaneous fitting to the spectrum in these two wavelength ranges 
with the $z \sim 2.203$ \lya\ and \lyb\ absorbers, and $z \sim 1.70$ contaminations, 
together with the other Ly$\alpha$ absorbers.
We obtain a self-consistent model that is shown with the red and blue curves in the left panels of Figure \ref{fig:X-shooterLLS}.

Based on the {\sc cloudy} model results, 
we find that System 2 is classified as a Lyman limit system (LLS), 
which is an optically thick clouds with an {\sc Hi} column density of $17.2 \leq \log N_\mathrm{H_I}$ (cm$^{-2}$) $\leq 20.3$. 
Note that the presence of this LLS does not change our conclusions. 
We confirm that the weak anti-correlation between \delg\ and \delF\ is found at the $\sim 80$~{\%} confidence level,
even if the LLS is masked out in the BOSS spectrum.
Our {\sc cloudy} model indicates that 
System 2 has a metallicity $Z/Z_\odot \simeq 0.02$ and an ionization parameter $\log U\simeq-3.0$ 
that are comparable with those of typical LLSs at $z\sim2$ \citep{Fumagalli2016a, Fumagalli2013a}.

Because Systems 1 and 3 have {\sc Hi} column densities of $\log N_\mathrm{H_I}$ (cm$^{-2}$) $\sim 16.0$ and $16.5$,
Systems 1 and 3 are classified as \lya\ forest absorbers
based on the moderately low column densities.
However, the {\sc cloudy} models imply that their metallicities are 
$Z/Z_\odot \simeq 0.3$ (System 1) and $1.0$ (System 3), 
which are two orders of magnitude higher than the median IGM metallicity \citep{Simcoe2011a}.  
These results would suggest that Systems 1 and 3 are gas clumps in the CGM and/or the intra-cluster medium of the proto-cluster candidate discussed in Section \ref{subsec:extreme value}.

\begin{deluxetable*}{cccccc}[]
\tablecolumns{6} 
\tablewidth{0pt} 
\tablecaption{X-shooter Observations\label{tab:observations}}
\tablehead{
    \colhead{Source}    & \colhead{R.A.} &  \colhead{Decl.} 
    & \colhead{Integration Time}  & \colhead{Dates of Observations}    & \colhead{S/N\tablenotemark{c}} \\
    \colhead{ } & \colhead{(J2000)} & \colhead{(J2000)} 
    & \colhead{(s)} & \colhead{ } & \colhead{[pix$^{-1}$]} 
}
\startdata 
COSMOS-QSO-199\tablenotemark{a} & $09\,58\,58.72$ & $+\,02\,01\,38.6$& UVB: 2700\tablenotemark{b} & 31 Dec 2010 & 27 \\ 
COSMOS-QSO-199\tablenotemark{a} & $09\,58\,58.72$ & $+\,02\,01\,38.6$&  VIS: 2700\tablenotemark{b} & 31 Dec 2010 & 20
\enddata 
\tablenotetext{$a$}{%
The background quasar of \Aobs, 
[VV2006] J095858.7+020138 at $z=2.448$
}
\tablenotetext{$b$}{%
Three individual exposures of 900 sec. 
}
\tablenotetext{$c$}{%
Median S/N per pixel for the combined spectra.
}
\end{deluxetable*} 
\begin{table*}[]
\begin{center}
\caption{
Column Densities of the Ions in the $z\sim2.203$ Absorption Systems 
\label{tab:fitting} }
\begin{tabular}{cccccc}
\hline

Ion & {\sc Cii} & Si{\sc iv} & {\sc Civ} & Mg{\sc ii}  \\ \hline
System 1 &&&& \\
$\log N$ (cm$^{-2}$) &  13.19 $\pm$ 0.70 &  13.14 $\pm$ 0.57 & 13.86 $\pm$ 0.19 & 12.31 $\pm$ 0.68 \\ 
\hline
System 2 &&&& \\
$\log N$ (cm$^{-2}$) &  13.92 $\pm$ 0.17 &  13.58 $\pm$ 0.24 & 14.29 $\pm$ 0.12 & 13.31 $\pm$ 0.21 \\
\hline
System 3 &&&& \\
$\log N$ (cm$^{-2}$) &  14.03 $\pm$ 0.16 &  13.54 $\pm$ 0.35 & 13.59 $\pm$ 0.35 & 13.29 $\pm$ 0.13  \\
\hline
\end{tabular}
\end{center}
\end{table*}

\section{C.  Counterparts for each extreme cylinder} \label{sec:APPENDIX C}
\begin{figure*}[h]
	\centering
	\includegraphics[clip,bb=0 00 500 520,width=0.6\hsize]{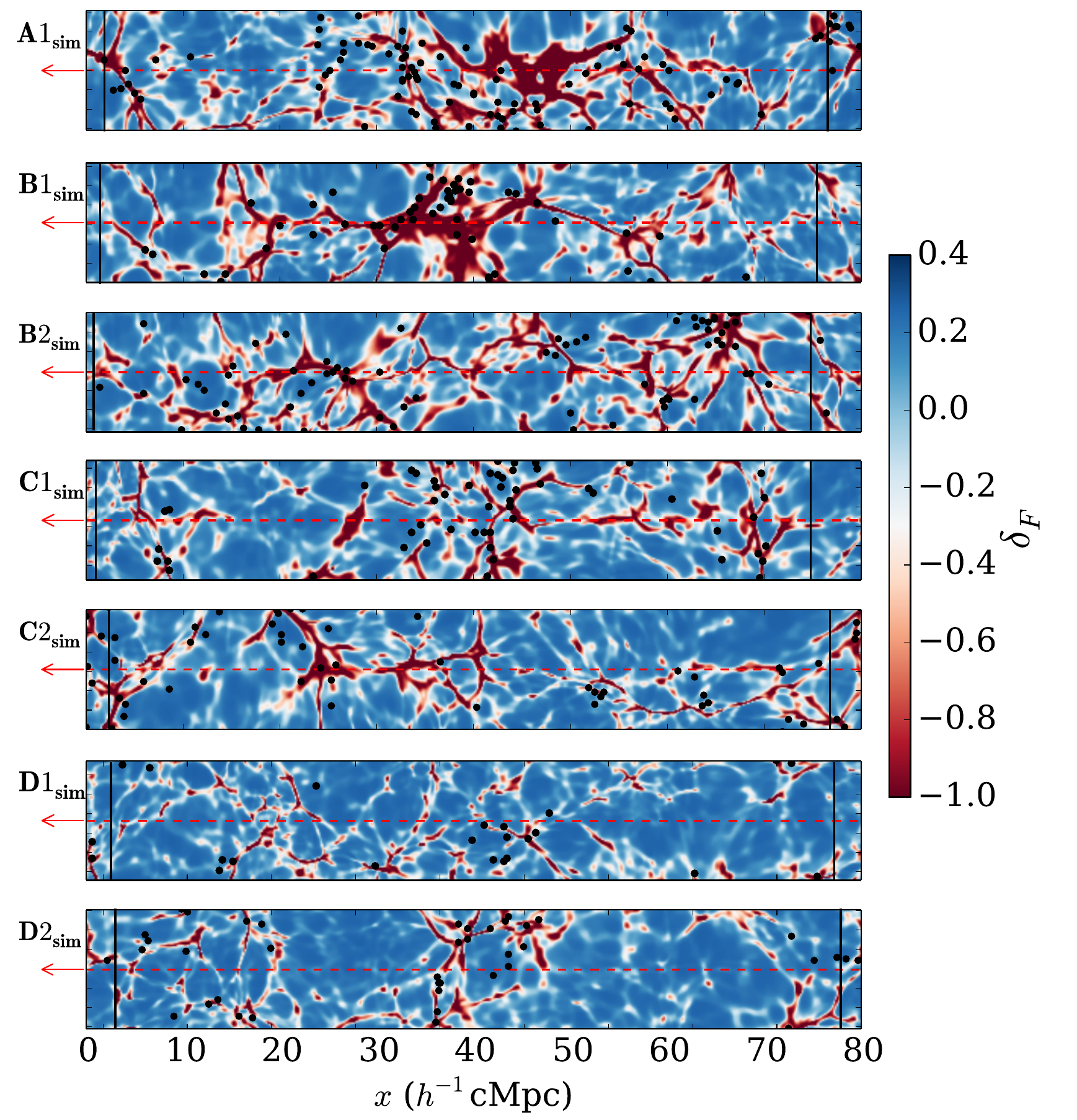}
	\caption{
	Same as Figure \ref{fig:simulation sightline}, but for ${\rm A1_{sim}}$,
	 ${\rm B1_{sim}}$,  ${\rm B2_{sim}}$, ${\rm C1_{sim}}$,  ${\rm C2_{sim}}$, 
	 ${\rm D1_{sim}}$, and ${\rm D2_{sim}}$.
	}
	\label{fig:counter part}  
\end{figure*}

In addition to \Asim - \Dsim, there exist counterparts of \Aobs-\Dobs\ 
in our simulations.
We find additional two counterparts for each extreme cylinder,
\Bobs, \Cobs, or \Dobs, in different volumes of the simulations,
which are referred to as (${\rm B1_{sim}}$,  ${\rm B2_{sim}}$), 
(${\rm C1_{sim}}$,  ${\rm C2_{sim}}$), or (${\rm D1_{sim}}$,  ${\rm D2_{sim}}$).
Moreover, we identify one additional counterpart of \Aobs, ${\rm A1_{sim}}$. We cannot find
another counterpart of \Aobs, because the large \delg\ and \delF\ values of \Aobs\ 
are very rare in the simulation box.
Figure $\ref{fig:overplot}$ presents the \delg\ and \delF\ values of 
these additional counterparts with the blue crosses.
These \delg -\delF\ values are comparable with 
those of \Asim - \Dsim\ at the $\sim 1\sigma$ error levels. 
Figure \ref{fig:counter part} shows that these sightlines penetrate 
large overdensities, gas filaments parallel with (orthogonal to) the sightline, or large voids.

\newpage
\bibliographystyle{apj}
\bibliography{citations}
\end{document}